\documentclass{iopart}

\pdfoutput=1

\usepackage{graphicx,epsfig,sidecap}
\usepackage{bm}

\def\be{\begin{equation}}
\def\ee{\end{equation}}

\def\bea{\begin{eqnarray}}
\def\eea{\end{eqnarray}}

\def\ffrac#1#2{\textstyle{#1\over#2}\displaystyle}

\def\Tr{{\rm Tr}}

\def\a{\alpha}

\gdef\zb{{\bar z}} \gdef\Db{{\overline\Delta}}

\begin{document}

\title[Entanglement entropy  of two disjoint intervals in CFT II] 
{Entanglement entropy of two disjoint intervals in conformal field theory II}

\author{Pasquale Calabrese$^1$, John Cardy$^{2}$, and Erik Tonni$^3$}
\address{$^1$Dipartimento di Fisica dell'Universit\`a di Pisa and INFN,
             Pisa, Italy.\\
         $^2$ Oxford University, Rudolf Peierls Centre for
          Theoretical Physics, 1 Keble Road, Oxford, OX1 3NP, United Kingdom
          and All Souls College, Oxford.\\
         $^3$ Center for Theoretical Physics, Massachusetts Institute of Technology, Cambridge, MA 02139, USA
          }

\date{\today}

\begin{abstract}

We continue the study of the entanglement entropy of two disjoint intervals in conformal field theories that we started 
in J.~Stat.~Mech.~(2009)~P11001. 
We compute $\Tr\rho_A^n$ for any integer $n$ for the Ising universality class and the final result is expressed as a sum
of Riemann-Siegel theta functions. These predictions are checked against existing numerical data.
We provide a systematic method  that gives the full asymptotic expansion of the 
scaling function for small four-point ratio (i.e.~short intervals). 
These formulas are compared with the direct expansion of the full results for free compactified boson and Ising model.
We finally provide the analytic continuation of the first term in this expansion in a completely analytic form.

\end{abstract}

\maketitle

\section{Introduction}

In this manuscript we continue the study of the entanglement entropies of two disjoint intervals in conformal field theories (CFT) that 
we started in Ref. \cite{cct-09}.
Let us consider a one-dimensional critical system whose scaling limit is described by a CFT of central charge $c$, 
and a partition into two parts $A$ and its complement $B$. 
We define the reduced density matrix $\rho_A=\Tr_B\rho$, where $\rho$ is the density matrix of the entire system.
The entanglement between $A$ and $B$ is characterized by the moments of $\rho_A$, i.e.~$\Tr\rho_A^n$, 
from which through the replica trick \cite{cc-04,cc-rev} one obtains the Von Neumann entanglement entropy \cite{revs}
\be
S_A\equiv-\Tr\rho_A\log \rho_A= -\left.\frac{\partial}{\partial n} \Tr\rho_A^n\right|_{n=1}\,.
\label{rt}
\ee
For integer $n$, $\Tr\rho_A^n$ is proportional to the partition function on an $n$-sheeted Riemann surface with branch cuts along 
the subsystem $A$, i.e.~$\Tr\rho_A^n=Z_n(A)/Z_1^n$ where $Z_n(A)$
is the partition function  of the CFT on a conifold where $n$ copies of the manifold ${\cal R}={\rm system}\times R^1$ (or ${\rm
system}\times S^1$ at finite temperature) are coupled along branch cuts along each connected piece of $A$ at a time-slice $t=0$.

In the case when $A$ is a single interval of length $\ell$ in an infinite system, the $n$-sheeted Riemann surface is 
topologically equivalent to the complex plane on which it can be mapped by a uniformizing mapping. 
By studying the transformation of the stress energy tensor (i.e.~the response of the partition function to a scale 
transformation) under this uniformizing map, one obtains \cite{cc-04,cc-rev}
\be
\Tr\rho_A^n=c_n{\ell}^{-\frac{c}6 (n-1/n)}\,,
\label{Trn}
\ee
from which the replica trick (\ref{rt}) leads to the famous result $(c/3)\log \ell$ \cite{Holzhey,Vidal}
and to the the full spectrum of $\rho_A$ \cite{cl-08}.

When the subsystem $A$ consists of $N$ disjoint intervals (always in an infinite system), 
the entanglement entropy is sensitive to the full operator content of the theory and not only to the central charge $c$. 
The $n$-sheeted Riemann surface ${\cal R}_{n,N}$ has genus $(n-1)(N-1)$ and cannot be mapped to the complex plane so that 
the CFT calculations become much more complicated. 
Only few results are analytically available. For two intervals, i.e.~$A=A_1\cup A_2=[u_1,v_1]\cup [u_2,v_2]$ (without loss of generality 
we assume $v_2>u_2>v_1>u_1$),
global conformal invariance leads to the scaling form
\be\fl
\Tr \rho_A^n\equiv \frac{Z_{\mathcal{R}_{n,2}}}{Z_1^n}
=c_n^2 \left(\frac{(u_2-u_1)(v_2-v_1)}{(v_1-u_1)(v_2-u_2)(v_2-u_1)(u_2-v_1)} 
\right)^{\frac{c}6(n-1/n)} {\cal F}_{n}(x)\,,
\label{Fn}
\ee 
where $x$ is the four-point ratio (for real $u_j$ and $v_j$, $x$ is real) 
\begin{equation}
x=\frac{(v_1-u_1)(v_2-u_2)}{(u_2-u_1)(v_2-v_1)}\,.
\label{4pR}
\end{equation}
In Ref. \cite{cct-09} we showed that for a boson compactified on a circle of radius $R$ (a Luttinger liquid field theory), the 
universal scaling function $\mathcal{F}_n(x)$ is
\begin{equation}
\mathcal{F}_n(x)=
\frac{\Theta\big({\bf 0}|\eta\Gamma\big)\,\Theta\big({\bf 0}|\Gamma/\eta\big)}{
[\Theta\big({\bf 0}|\Gamma\big)]^2}\,,
\label{Fnv}
\end{equation}
where $\Gamma$ is an $(n-1)\times(n-1)$ matrix with elements
\be
\Gamma_{rs} =  
\frac{2i}{n} \sum_{k\,=\,1}^{n-1} 
\sin\left(\pi\frac{k}{n}\right)\beta_{k/n}\cos\left[2\pi\frac{k}{n}(r-s)
\right]\,, 
\label{Gammadef}
\ee
and 
\be
\beta_y=\frac{\, _2 F_1(y,1-y;1;1-x)}{\, _2 F_1(y,1-y;1;x)}\,.
\label{betadef}
\ee
$\eta$ is a universal critical exponent proportional to the square of the 
compactification radius $R$, while $\Theta$ is the Riemann-Siegel 
theta function
\begin{equation}
\label{theta Riemann def}
\Theta({\bf z}|\Gamma)\,\equiv\,
\sum_{\mathbf{m} \,\in\,\mathbf{Z}^{n-1}}
\exp\big[\,i\pi\,\mathbf{m}^{\rm t}\cdot \Gamma \cdot {\bf m}+2\pi i {\bf m}^{\rm t}\cdot {\bf z}\,\big]\,,
\end{equation}
with ${\bf z}$ a generic complex vector with $n-1$ components (we denote vectors in bold letter to  
easily distinguish them from scalars).
For $n=2$, $\mathcal{F}_2(x)$ reduces to the result in \cite{fps-09}.
This is the only result for  $\mathcal{F}_n(x)$ known in full generality, and 
many other important partial results are known \cite{fps-09,cg-08,ch-04,ffip-08,kl-08,rt-06,hr-08,atc-09,ip-09,fc-10,h-10,fc-10b,t-10,c-10,Neg}
both analytically and numerically. 

However, although the moments of the reduced density matrix have been obtained for all integer $n$, the analytic continuation to
complex $n$ is still beyond our knowledge and so is the Von Neumann entropy. 
Some results have been obtained in Ref. \cite{cct-09} from Eq. (\ref{Fnv}) in the limit of small $\eta$ and of small $x$ 
(the latter only numerically, except for few specific values of $\eta$), and the two limits do not commute.

In this manuscript we extend the known results in several directions. 
We provide a systematic method that gives the full asymptotic expansion of ${\cal F}_n(x)$ for small  $x$
(i.e.~short intervals). This method is valid for any conformal invariant theory (even in the presence of boundaries and at finite 
temperature) and the obtained results are tested against known results (Sec.~\ref{Secope}).
Furthermore, it provides the coefficients of any expansion when the lengths of the intervals are smaller than any other length of 
the system, such as finite size, inverse temperature etc. 
In these cases the entanglement  entropy is sensitive to all the CFT data on the appropriate Riemann surface. 
We derive the scaling function ${\cal F}_n(x)$ for all $n$ for the Ising universality class, generalizing the result in Ref. \cite{atc-09} 
for $n=2$ (in Sec.~\ref{SecIM}).
We extend the small $x$ expansion to the second non vanishing order both for the compactified boson and for the Ising model 
(Sec.~\ref{Secsx}). 
We derive a full analytic formula for the first non vanishing term in the small $x$ expansion of the Von Neumann entropy 
(Sec.~\ref{ancont}). 

\section{Summary of results}

This section contains a summary of our main results, that are derived in the following.

\subsection{Short-length expansion}

We develop a general framework to calculate the {\it short-length expansion} of $\Tr\rho_A^n$ 
that is very similar to the standard short-distance expansion for correlators. 
This formalism (that extends and generalizes a recent approach by Headrick \cite{h-10}) 
applies every time in which $A$ consists of one or more intervals $\{I_p\}$
whose lengths $\ell_p$ are much smaller than all the other characteristic lengths $\sim L$ (for example, the separation
between the intervals, the size of the system or the inverse temperature). 
The short length expansion provides a series of powers of $\ell_p/L$ whose terms are universal and encode {\it all} the data 
of the CFT such as the scaling dimensions and the operator product expansion (OPE) coefficients.
This expansion makes manifest the common belief that the entanglement entropies of disjoint intervals encode
information about the {\it full data} of the CFT.

When these arguments are specialized to the case of two intervals (of lengths $\ell_1$ and $\ell_2$ whose centers are at a 
distance $r$) in an infinite system, we obtain
\be
{\rm Tr}\,\rho_A^n=
\frac{c_n^2}{(\ell_1\ell_2)^{c/6(n-1/n)}} \sum_{\{k_{j}\}}  \left(\frac{\ell_1\ell_2}{n^2 r^2}\right)^{\sum_j (\Delta_j+\Db_j)} s_{\{k_j\}}(n)\,,
\ee
where the sum is over all the set of possible operators $\{\phi_{k_j}\}$ of the CFT with conformal dimensions 
($\Delta_{k_j}, \Db_{k_j}$). 
$s_{\{k_j\}}(n)$ are calculable coefficients depending on the correlation functions of these operators.
This is easily converted in a systematic small $x$ expansion of the function ${\cal F}_n(x)$.
Each set of operators ${\{k_j\}}$ gives a leading contribution at order $x^{\sum_j (\Delta_j+\Db_j)}$. 
Furthermore, in ${\cal F}_{n}(x)$
there are analytic contributions in $x$, which do not correspond to any operator and they come 
expanding the prefactor in Eq. (\ref{Fn}) in powers of $x$. 

In the case of compactified free boson and Ising model, we can expand the exact results for any $x$ to obtain
\be
{\cal F}_n(x)=1+ \left(\frac{x}{4n^2}\right)^{ \a}s_2(n)+ \left(\frac{x}{4n^2}\right)^{2\a}s_4(n)+\dots \,,
\label{Fexpintro}
\ee
where $\a=\min[\eta,1/\eta]$ for boson and $\a=1/4$ for Ising. 
The coefficients $s_2(n)$ and $s_4(n)$ are explicitly calculated in the asymptotic expansion and they are identified with 
the contributions of the short-length expansion coming from the two- and four-point functions of the most relevant operator in each 
theory. Contributions from three-point function are subleading. 
$s_2(n)$ has a simple zero at $n=1$, while $s_4(n)$ has simple zeros at $n=1,2,3$.
In general $s_m(n)$ (coming from an $m$-point function) will have simple zeros at any integer $n<m$.

\subsection{Result for the Ising universality class}
For the Ising model, the scaling function ${\cal F}_n (x)$ is 
\begin{equation}
{\cal F}_n (x)= 
\frac{1}{2^{n-1}\Theta({\bf 0}|\Gamma)} \sum_{\bm{\varepsilon,\delta}}
\left| \Theta\bigg[\begin{array}{c} \bm{\varepsilon} \\ \bm{\delta}  \end{array}\bigg] ({\bf 0}|\Gamma)\right|\,.
\label{Fn ising}
\end{equation}
Here $\Theta$ is the Riemann theta function with characteristic defined as
\begin{equation}\fl
\label{def theta}
\Theta
\bigg[\begin{array}{c} \bm{\varepsilon} \\ \bm{\delta}  \end{array}\bigg] ({\bf z} |\Gamma)
\,\equiv\,
\sum_{{\bf m} \in \mathbf{Z}^g} \exp\Big[i \pi ({\bf m+{\bm \varepsilon}})^{{\rm t}}\cdot \Gamma\cdot ({\bf m+\bm{\varepsilon}})
+2\pi i\,({\bf m+\bm{\varepsilon}})^{{\rm t}}\cdot ({\bf z+\bm{\delta}})\Big]\,,
\end{equation}
where ${\bf z} \in \mathbf{C}^{n-1}$ and $\Gamma$ is the same as in Eq. (\ref{Gammadef}).
$\bm{\varepsilon}, \bm{\delta} $ are vector with entries  $0$ and $1/2$.
The sum in $(\bm{\varepsilon,\delta})$ in Eq. (\ref{Fn ising}) is intended over all the $2^{n-1}$ vectors 
${\bm \varepsilon}$ and ${\bm \delta}$ with these entries. 
We show that ${\cal F}_n (x)={\cal F}_n (1-x)$ and that it reproduces known analytical and numerical results \cite{atc-09,fc-10}.

\subsection{Analytic continuation of the leading term in the small $x$ expansion}

From the short length expansions (and from the asymptotic expansion of known results), the leading term of the 
small $x$ expansion in ${\cal F}_n(x)$ is \be
s_2(n)={\cal N}\, \frac{n}2
\sum_{j=1}^{n-1}
\frac1{\left[ \sin\left(\pi\frac{j}{n}\right) \right]^{2\alpha}}\,,
\ee
and it comes from the two-point function of the most relevant operator of the CFT.
The integer ${\cal N}$ counts the number of inequivalent correlation functions giving the same contribution (coming most often 
from degenerate operators).
We have ${\cal N}=2$ for free boson and ${\cal N}=1$ for the Ising model.

We provide the analytic continuation to complex $n$ for any value of $\a$ whose derivative at $n=1$ is
\be
s'_2(1)= {\cal N} \frac{\sqrt{\pi}\,  \Gamma (\a+1)}{4 \Gamma \left(\a+\frac{3}{2}\right)}\,,
\label{s21fin}
\ee
that gives the asymptotic expansion of the scaling function of the von Neumann entropy 
\be
{\cal F}_{VN}(x)\equiv \frac{\partial}{\partial n} {\cal F}_n (x)\Big|_{n=1}=
{\cal N} \left(\frac{x}4\right)^{\a}  \frac{\sqrt{\pi}\, \Gamma (\a+1)}{4 \Gamma \left(\a+\frac{3}{2}\right)}+\cdots 
\,.
\ee

\section{Short interval expansion}
\label{Secope}

In this section we consider $\Tr\rho_A^n$ in a variety of situations in which $A$ consists of one or more intervals $\{I_p\}$
whose lengths $\ell_p$ are much smaller than all the other characteristic lengths $\sim L$ (for example, the separation
between the intervals, the size of the system or the inverse temperature) and show that it can be expanded in a series of
powers of $\ell_p/L$ whose terms are universal and encode all the data of the CFT such as the scaling
dimensions and the operator product expansion (OPE) coefficients.
Our arguments are a generalization of those of Headrick \cite{h-10} to the case of non-zero conformal spin, to the 
next to leading orders in the OPE and to the case of more than two intervals. 
When these arguments are specialized to the case $N=2$ in an infinite system, they provide a systematic small $x$
expansion of the function ${\cal F}_n(x)$.

The basic idea, familiar from other situations in CFT, is that this ratio of
partition functions can be thought of as a correlator
\be
\label{eq:intervals}
\frac{Z_n(\{I_p\})}{Z_1^n}=\langle\prod_{p=1}^N I_p\rangle_{{\cal R}^n}\,,
\ee
in $n$ copies of the CFT on \em decoupled \em manifolds $\cal R$ (notice $n$ as power of ${\cal R}$ and not as 
a subscript). 
More explicitly, in this correlator each $I_p$ may be expanded in
a complete set of local fields at a given point in the interval,
for example its midpoint $z_p$:\footnote{This may be thought of as a modification of the standard OPE in the plane 
$$
\phi_a(u)\phi_b(v)=
\sum_c C_{abc}\, (u-v)^{-\Delta_a-\Delta_b+\Delta_c} (\bar{u}-\bar{v})^{-\Db_a-\Db_b+\Db_c} \phi_c \big(\ffrac{u+v}2\big)\,,
$$
to the case when when have non-local twist operators in $n$ copies of the CFT.
}
\begin{equation}\label{eq:Ip}
I_p=\sum_{\{k_j\}}C_{\{k_j\}}\prod_{j=1}^n\phi_{k_j}(z_{p_j})\,,
\end{equation}
where $\{\phi_k\}$ denotes a complete set of local fields for a
single copy of the CFT, and $z_{p_j}$ is the point $z_p$ on the $j$th sheet.

This statement may be understood most easily within radial quantization. In the limit when the lengths $\ell_p$ are small
compared with the separation between the intervals (and other length scales such as the system size, the distance to any
boundary or the inverse temperature), we may surround the interval $I_p$ by a circle on each sheet and consider the state
$|I_p\rangle$ induced by coupling the CFTs along the branch cut $I_p$. 
Since at this radius the $n$ sheets are distinct, this state lies in the Hilbert space $\otimes_{j=1}^n{\cal H}_j$ where
${\cal H}_j$ is the space of the $j$th CFT in radial quantization about $z_{p_j}$. Each of these is spanned by states
$|\phi_k\rangle$ in one-to-one correspondence with the scaling operators of the CFT. That is, we can write
\be
|I_p\rangle=\sum_{\{k_j\}}C_{\{k_j\}}\otimes_j|\phi_{k_j}\rangle_j\,.
\ee
However, by the operator-state correspondence of CFT, this is equivalent to Eq. (\ref{eq:Ip}).

The main point is now that, like the more conventional short-distance expansion of two local operators, the coefficients
$C_{\{k_j\}}$ for a given interval are independent of the location and lengths of the other intervals in Eq. (\ref{eq:intervals}).
Therefore they may be determined by considering the simplest possible case, that of a single interval $I=(-\ell/2,\ell/2)$ on
the infinite line. In this case the partition function itself is trivial, but we can also consider the insertion of other operators
${\cal O}_j$ on each sheet outside the circle of radial quantization. Specifically
\be\fl
\frac{Z_n(A)}{Z_1^n}\langle\prod_j{\cal O}_j\rangle_{{\cal R}_{n,1}}=
\langle I\,\prod_j{\cal O}_j\rangle_{{\cal R}^n} =
\sum_{\{k_j\}}C_{\{k_j\}}\prod_j\langle\phi_{k_j}(0_j){\cal O}_j\rangle_{{\cal R}_j}\,.
\ee
Note the appearance of the ratio of partition functions, reflecting the fact that the correlators on the left and right
sides are evaluated in different geometries. ${\cal R}_j$ is the $j$th copy of the complex plane.

In particular, by choosing the ${\cal O}_j$ to be a complete set
of operators $\phi_{k_j}(z_j)$ as $z_j\to\infty$, we may use
orthogonality\footnote{This assumes that the operators $\phi_k$
are all quasiprimary, that is $L_1\phi_k=0$.}
\be
\langle\phi_k(0)\phi_{k'}(z)\rangle=
z^{-2\Delta_k}\zb^{-2\Db_k}\,\delta_{kk'}\,,
\ee
where $(\Delta_k,\Db_k)$ are the conformal dimensions of $\phi_k$, to find that
\begin{equation}\label{eq:Ck}
C_{\{k_j\}}=\frac{Z_n}{Z_1^n}\lim_{z_j\to\infty_j}
z^{2\sum_j\Delta_{k_j}}\zb^{2\sum_j\Db_{k_j}}
\langle\prod_j\phi_{k_j}(z_j)\rangle_{{\cal R}_{n,1}}\,,
\end{equation}
where the correlation function on the right is evaluated on the
$n$-sheeted surface corresponding to a single interval in an
infinite system. Note that this correlation function vanishes
unless the fusion of $\prod_j\phi_{k_j}$ includes the conformal
block of the identity. Also their total conformal spin should be
an even integer, since $I_p$ is invariant under a rotation through
$\pi$ about its midpoint.

We normalize the single interval partition function according to Eq. (\ref{Trn}), so that
\be
C_{\{k_j\}}=c_n \ell^{-(c/6)(n-1/n)+\sum_j(\Delta_{k_j}+\Db_{k_j})}\,d_{\{k_j\}}\,,
\label{dvsC}
\ee
where the $d_{\{k_j\}}$ are dimensionless universal numbers.

In the case where the $\phi_{k_j}$ are all \em primary\em, the
correlator in Eq. (\ref{eq:Ck}) is simply related to the same
correlator in the plane by the uniformizing map
\be
z\to
f(z)=\left(\frac{z-\ffrac12\ell}{z+\ffrac12\ell}\right)^{1/n}\,,
\ee
which maps each sheet ${\cal R}_j$ in a wedge of angle $2\pi/n$.
Infinity on the $j$th sheet is mapped to the $n$th root of unity $e^{2\pi ij/n}$, and $f'(z)\sim(\ell/nz^2)$. 
Thus, for primary operators,
\be
\langle\prod_j\phi_{k_j}(z_j)\rangle_{{\cal R}_{n,1}}=
\prod_jf'(z_j)^{\Delta_{k_j}}{\overline{f'(z_j)}}^{\Db_{k_j}}
\langle\prod_j\phi_{k_j}\big(f(z_j)\big)\rangle_{\bf C}\,.
\label{Clast}
\ee
So, using Eqs. (\ref{eq:Ck}), (\ref{dvsC}), and (\ref{Clast}), we have
\be
d_{\{k_j\}}=n^{-\sum_j(\Delta_{k_j}+\Db_{k_j})}
\langle\prod_{j=1}^n\phi_{k_j}\big(e^{2\pi ij/n}\big)\rangle_{\bf C}\,.
\label{main}
\ee
For non-primary operators, derivatives and inhomogeneous terms
involving the central charge may also arise. 
Eq. (\ref{main}) is the key result of this section from which all the others follow.

In the limit $\ell\to0$, the sum over the other values of $\{k_j\}$ may be organized according to how many of them are non-zero
(at least in a unitary CFT with non-negative scaling dimensions).
\begin{itemize}
\item The leading term as $\ell\to0$ is given by taking all the $\Delta_{k_j}=0$, that is $\phi_{k_j}={\bf 1}$, the identity operator. 
\item 
%
%
The next term comes from taking all the $k_j=0$ except one, say $k_j=k$. 
However in this case $\phi_k$ cannot be primary, since the 1-point function $\langle\phi_k\rangle_{\bf C}$ would vanish.
The most interesting case is when $\phi_k$ is a component of the stress tensor, $T$ or $\overline T$. 
In that case Eq. (\ref{eq:Ck}) becomes
\be
C_T=C_{0,\ldots,T,\ldots,0}=\frac{Z_n}{Z_1^n}\frac2c \lim_{z_j\to\infty}z_j^4\langle T(z_j)\rangle_{{\cal R}_{n,1}}\,.
\ee
(Note the factor of $2/c$ because the 2-point function of $T$ is normalized to $c/2$ rather than unity.) However \cite{cc-04}
\be
\langle T(z)\rangle_{{\cal R}_{n,1}}=\frac{c}{12}\frac{(1-1/n^2)}
{(z-\ffrac12\ell)^2(z+\ffrac12\ell)^2}\,,
\ee
so that
\be
d_T=\frac16\left(1-\frac1{n^2}\right)\,,
\ee
and similarly for $d_{\overline T}$.
\item 
The next contribution comes from taking {\it two} of the $k_j$ to be non-zero. 
Since the product must couple to the identity block this requires the two operators to be in the same block, and if they
are real and primary they must be the same operator. 
Thus, taking $k_{j_1}=k_{j_2}=k$, we have, in an obvious notation,
\begin{eqnarray}\fl
d_{k}^{(j_1j_2)}&=&n^{-2(\Delta_k+\Db_k)}\langle\phi_k(e^{2\pi
ij_1/n})\phi_k(e^{2\pi ij_2/n})\rangle_{\bf C}\nonumber\\ \fl
&=&\frac{n^{-2(\Delta_k+\Db_k)}}{\big[e^{2\pi ij_1/n}-e^{2\pi ij_2/n}\big]^{2\Delta_k}
\big[e^{-2\pi ij_1/n}-e^{-2\pi ij_2/n}\big]^{2\Db_k}}\label{eq:dkjj}\,.
\end{eqnarray}
Note that we must have $j_1\not=j_2$. 
\item
Similarly, if {\it three} of the $k_j$ are non-zero and correspond to primary fields $(\phi_{k_1},\phi_{k_2},\phi_{k_3})$
\begin{eqnarray}\label{d3}
\fl
&&d_{k_1k_2 k_3}^{(j_1j_2j_3)}=
n^{-\Delta_{1}-\Delta_{2}-\Delta_{3}-\Db_{1}-\Db_{2}-\Db_{3}}
\langle\phi_{k_1}(e^{\frac{2\pi ij_1}n})\phi_{k_2}(e^{\frac{2\pi ij_2}n})\phi_{k_3}(e^{\frac{2\pi ij_3}n})\rangle_{\bf C}\\ \fl
&&=\frac{n^{-\Delta_{1}-\Delta_{2}-\Delta_{3}-\Db_{1}-\Db_{2}-\Db_{3}}
c_{123}}{
[e^{\frac{2\pi ij_1}{n}}-e^{\frac{2\pi ij_2}n}]^{\Delta_{1}+\Delta_{2}-\Delta_{3}}
[e^{-\frac{2\pi ij_1}n}-e^{-\frac{2\pi ij_2}n}]^{\Db_{1}+\Db_{2}-\Db_{3}}\times{\rm
cyclic\ permutations}},\nonumber
\end{eqnarray}
where we shortened $\Delta_{k_p}=\Delta_p$ and $c_{123}=c_{k_1k_2k_3}$ is an OPE coefficient. 
\item
The next term involves the 4-point functions at the $n$th roots of unity, and so on.
However, the correlation functions with more than 3 points are different for any CFT  and so Eq. (\ref{main}) 
cannot be simplified in general.
\end{itemize}

Note that for a fixed positive integer $n$, only the
$N$-point functions with $N\leq n$ contribute. This means that the
coefficients in the short length expansion, as analytic functions
of $n$, must have zeroes at positive integer values of $n<N$.

We now discuss a number of applications of the short length
expansion.

\subsection{Single interval in a finite geometry or near a boundary.}

In this case the 1-point functions $\langle\phi_{k_j}\rangle_{{\cal R}_1}$ can be non vanishing. 
An interesting example is the contribution of the stress tensor. From the above we have
\be\fl
{\rm
Tr}\,\rho_A^n=c_n\ell^{-(c/6)(n-1/n)}\left(1+\cdots+\ffrac16(n-1/n)\ell^2(\langle
T\rangle+\langle\overline T\rangle)+\cdots\right)\,.
\ee
(The extra factor of $n$ comes from the sum over $j$.) We
recognize the second factor as being proportional to the energy
density $T_{tt}$. In fact we can deduce a rather general result:
the $O(\ell^2)$ correction to the R\'enyi entropy $S_n=\frac1{1-n} \ln \tr\rho_A^n$ of a single
interval in a finite system (or at finite temperature) is given by
the local energy density:
\begin{equation}\label{eq:deltaS}
\delta S_n=\frac\pi 6(1+1/n)\langle{\cal E}\rangle\,\ell^2\,.
\end{equation}
This can be checked by comparison with the result in
Ref.~\cite{cc-04} for a single interval in a system of length $L$
with periodic boundary conditions
\be
S_n=\frac{c}6\left(1+\frac1n\right) \log\left[\frac{L}\pi\sin \frac{\pi\ell}L\right]\,,
\ee
by expanding to $O((\ell/L)^2)$ and using $\langle{\cal E}\rangle=-\pi c/6L^2$. However Eq. (\ref{eq:deltaS}) is more general,
and applies, for example, to the case of an interval in the
interior of an finite open system (as considered in Ref. \cite{fc-10}), or to a finite system at finite
temperature. In fact at finite temperature we see that this
correction (which is valid for $\ell\ll\beta$) already exhibits
the beginning of the crossover to thermodynamic entropy which
actually occurs only in the limit $\ell\gg\beta$ (see Ref. \cite{cc-04}).

Another case where 1-point functions can arise is in the case when
the interval is near a boundary (say, at a distance $\sim y$) with
$\ell\ll y$. In that case the leading correction comes from the
case where two of the $\{k_j\}$ are non-zero, equal to (say) $k$,
and it involves the square of the 1-point function
$\langle\phi_k(y)\rangle=f_k\,y^{-\Delta_k-\Db_k}$, where $f_k$ is
universal (but depends on the boundary conditions). The correction
to ${\rm Tr}\,\rho_A^n$ is then
\be
\frac12\sum_{j_1\not=j_2}d_{k}^{(j_1j_2)}\,f_k^2\left(\frac\ell{y}\right)^{2\Delta_k+2\Db_k}\,.
\ee

\subsection{Two intervals in an infinite system.}

We now apply the general result to the case of interest for the rest of this paper, i.e.~we take 
$A=[u_1,v_1]\cup[u_2,v_2]$. Global conformal invariance gives  $\Tr\rho_A^n$ as in
Eq. (\ref{Fn}) with the harmonic ratio in Eq. (\ref{4pR}).
In terms of the notation above, we have $\ell_1=|u_1-v_1|$ and $\ell_2=|u_2-v_2|$. 
Let $r=\frac12|u_1+v_1-u_2-v_2|$ be the distance between the centers of the intervals. 
We are interested in the limit where $\ell_1,\ell_2\ll r$. 
In that case 
\be
x=\frac{\ell_1\ell_2}{r^2-\left(\ffrac{\ell_1-\ell_2}2\right)^2}=
\frac{\ell_1\ell_2}{r^2}\left[1+ O\left(\left(\ell_p/{r}\right)^2\right)\right]\,, 
\label{xlarr}
\ee
and the prefactor in Eq. (\ref{Fn}) is 
\be\fl
\frac{(u_2-u_1)(v_2-v_1)}{(v_1-u_1)(v_2-u_2)(v_2-u_1)(u_2-v_1) }=
\frac1{\ell_1\ell_2(1-x)}=
\frac1{\ell_1\ell_2}\big(1+x+O(x^2)\big)\,.
\label{pref}
\ee
Notice that often $r$ is defined as the distance between the two intervals $|u_2-v_1|$, but the current choice absorbs some of 
the subleading corrections as clear from the fact that Eq. (\ref{xlarr}) has no corrections at order $1/r$ that would be present with any 
other choice of $r$. Furthermore if $\ell_1=\ell_2$, we have $x=\ell_1^2/r^2$ exactly.

From Eq. (\ref{eq:intervals}), we  have
\be
{\rm Tr}\,\rho_A^n=
\langle I_1I_2\rangle_{{\bf C}^n}\,,
\ee
and inserting the short interval
expansion (\ref{eq:Ip}) for each interval, we find
\bea
{\rm Tr}\,\rho_A^n&=&\sum_{\{k_{1_j}\}}\sum_{\{k_{2_j}\}}C_{\{k_{1_j}\}}(\ell_1)
C_{\{k_{2_j}\}}(\ell_2)\prod_{j=1}^n\langle
\phi_{k_{1_j}}(r)\phi_{k_{2_j}}(0)\rangle_{\bf C} \nonumber \\ &=&
\sum_{\{k_{j}\}}C_{\{k_{j}\}}(\ell_1)
C_{\{k_{j}\}}(\ell_2)\prod_{j=1}^n\langle
\phi_{k_{j}}(r)\phi_{k_{j}}(0)\rangle_{\bf C}\,,
\eea
where in the second line we used orthogonality, i.e.~the only non-zero contributions are
from $\{k_{j}\}\equiv\{k_{1_j}\}=\{k_{2_j}\}$.\footnote{This assumes that the whole
system is critical and described by the CFT. In fact the short
interval expansion can also be used in the non-critical theory, as
long as $\ell_p\ll$ the correlation length $\xi$, in which case
these correlators are proportional to the Zamolodchikov metric.}
Using now Eqs. (\ref{dvsC}) and (\ref{main}) we have
\bea\fl
{\rm Tr}\,\rho_A^n&=& c_n^2 (\ell_1\ell_2)^{-c/6(n-1/n)} \sum_{\{k_{j}\}} (\ell_1\ell_2)^{\sum_j (\Delta_j+\Db_j)} d_{\{k_j\}}^2
\prod_{j=1}^n\langle \phi_{k_{j}}(r)\phi_{k_{j}}(0)\rangle_{\bf C} \nonumber\\ \fl &=&
c_n^2 (\ell_1\ell_2)^{-c/6(n-1/n)} \sum_{\{k_{j}\}}  \left(\frac{\ell_1\ell_2}{r^2}\right)^{\sum_j (\Delta_j+\Db_j)} d_{\{k_j\}}^2
\nonumber\\ \fl &=&
c_n^2 (\ell_1\ell_2)^{-c/6(n-1/n)} \sum_{\{k_{j}\}}  \left(\frac{\ell_1\ell_2}{n^2 r^2}\right)^{\sum_j (\Delta_j+\Db_j)}
\langle\prod_{j=1}^n\phi_{k_j}\big(e^{2\pi ij/n}\big)\rangle_{\bf C}^2\,.
\eea
Thus we obtain an expansion of ${\rm Tr}\,\rho_A^n$ in powers of $\ell_1\ell_2/r^2$, 
which, via Eq. (\ref{xlarr}), can be converted to the expansion of the function ${\cal F}_n(x)$ in powers of $x$.

As explained above, to organize the sums as an asymptotic expansion (i.e.~with increasing powers of $x$) we have to order all possible 
$n$-tuples of quasi-primaries $\phi_{k_j}$ with non vanishing correlation functions according to the value of $\sum_j (\Delta_j+\Db_j)$. 

\subsubsection{The contribution of the stress-energy tensor.}
There is an important term when one of the $\phi_{k_j}=T$ (or $\overline T$) and the rest of the $k_j=0$. 
Using the above values of $d_T$ and the fact that $\langle T(r)T(0)\rangle=c/(2r^4)$, this gives a contribution
\be
2\frac{cn(1-1/n^2)^2}{72}\left(\frac{\ell_1\ell_2}{r^2}\right)^2= \frac{cn(1-1/n^2)^2}{36} x^2\,,
\ee
where the factor $2$ in front of the lhs takes into account that there are two identical contributions from $T$ and $\overline T$
correlations.
Thus, in the infinite line,  there are no contributions at order $O(x)$ in $\Tr\rho_A^n$ coming from one-point functions.
There could also be terms $O(\ell^2)$ arising from two fields each of dimension $1$.
\subsubsection{The leading contribution.}
The leading term, assuming that there are operators of sufficiently low dimension, comes from the case where {\it two} of the
$\{k_j\}$ are non-zero, say equal to $k$. 
Their contribution in ${\cal F}_n(x)$ is
\begin{equation}
\label{eq:2ptcorr}
x^{2(\Delta_k+\Db_k)}\sum_{j_1<j_2} (d_{k}^{(j_1j_2)})^2\,,
\end{equation}
where $d_{k}^{(j_1j_2)}$ is given by Eq. (\ref{eq:dkjj}). 
There is an important point  to be made here concerning the role of
conformal spin $s_k=\Delta_k-\Db_k$. 
The expression for $d_{k}^{(j_1j_2)}$ is periodic under $(j_1,j_2)\to(j_1+n,j_2+n)$, and therefore
we can shift the summation variables $(j_1,j_2)\to(j_1+1,j_2+1)$ in Eq. 
(\ref{eq:2ptcorr}) without changing the result. However
\be
(d_{k}^{(j_1+1,j_2+1)})^2=e^{-8\pi i(\Delta_k-\Db_k)/n}\,(d_{k}^{(j_1j_2)})^2\,.
\ee
Hence the sum in Eq. (\ref{eq:2ptcorr}) vanishes unless $4s_k/n$ is an integer. 
This means that the contribution of a given primary field
with non-zero spin is non vanishing only for a finite number of values of $n$ (which are multiples of 4).

From above, we can easily write down the coefficient $s_k(n)$ entering in the expansion  of the function ${\cal F}_n(x)$ in power of $x$
as 
\bea
s_k(n)&=& 2^{4x_k} \sum_{0\leq j_1<j_2\leq n-1}
\frac{e^{4\pi i (j_1+j_2)s_k/n}}{|e^{2\pi i j_1/n}- e^{2\pi i j_2/n}|^{4x_k}}
\nonumber\\& =&
 \sum_{0\leq j_1<j_2\leq n-1}
\frac{e^{4\pi i (j_1+j_2)s_k/n}}{|\sin(\pi (j_2-j_1)/n)|^{4x_k}}\,,
\eea
where we defined $x_k=\Delta_k+\Db_k$. 
Since $4s_k/n$ is an integer,  the numerator is at most a sign.
For $s_k=0$, using the translational invariance in $j_{1,2}$, the sum can be rewritten as
\be
s_k(n)=\frac{n}2 \sum_{j=1}^{n-1}
\frac1{(\sin \pi j/n)^{4x_k}}\,, 
\label{s2sle}
\ee
that reproduces the result in \cite{h-10}.
In any specific CFT, this contribution should be multiplied by the number of inequivalent correlation functions giving 
identical contributions (due e.g.~to degenerate operators).

\subsubsection{The contribution of the three-point functions of primary operators.}
In the case of a three point function, the contribution in ${\cal F}_n(x)$ is given by
\begin{equation}
\label{eq:3ptcorr}
x^{
\Delta_{k_1}+\Db_{k_1}+\Delta_{k_2}+\Db_{k_2}+\Delta_{k_3}+\Db_{k_3}}
\sum_{0\leq j_1<j_2<j_3\leq n-1}(d_{k_1k_2k_3}^{(j_1j_2j_3)})^2\,,
\end{equation}
where $d_{k_1k_2k_3}^{(j_1j_2j_3)}$ is given by Eq. (\ref{d3}). 
Analogously to before, the contribution of non-zero spin operators vanishes unless
$2(s_{k_1}+s_{k_2}+s_{k_3})/n$ is an integer.
We recall that in the following equations, to have a non vanishing contribution, we also need the coefficient $c_{123}$
to be non-zero (for example for the Ising model, the three point functions of the spin and energy operator are both vanishing, 
and the one with lower scaling dimension would be $\langle \sigma(z_1) \sigma(z_2)\epsilon(z_3)\rangle$).
Thus, in most of the physical relevant cases, the contributions from three points functions are subleading compared to 
four-point ones, as we shall see in specific examples.

Specializing to the case of zero conformal spin, we have the coefficient 
entering in the expansion of the function ${\cal F}_n(x)$ in powers of $x$
$s_{123}(n)\equiv s_{k_1k_2k_3}(n)$: 
\bea\fl
s_{123}(n)&=&  \sum_{j_1<j_2<j_3}
\frac{2^{2(x_1+x_2+x_3)}c_{123}}{|e^{\frac{2\pi ij_1}{n}}-e^{\frac{2\pi ij_2}n}|^{2x_{12;3}}
|e^{\frac{2\pi ij_1}{n}}-e^{\frac{2\pi ij_3}n}|^{2x_{13;2}}
|e^{\frac{2\pi ij_2}{n}}-e^{\frac{2\pi ij_3}n}|^{2x_{23;1}}}\nonumber\\\fl
&=&
\sum_{0\leq j_1<j_2<j_3\leq n-1}\frac{c_{123}}{
[\sin(\pi j_{21}/n)]^{2x_{12;3}} [\sin(\pi j_{31}/n)]^{2x_{13;2}}
[\sin(\pi j_{32}/n)]^{2x_{23;1}}}\,,
\eea
where $j_{pq}=j_p-j_q$ and $x_{ab;c}=x_a+x_b-x_c$. Using the translational invariance in the $j_p$, one of the three sums 
can be removed, but doing so is not very illuminating. 

\subsubsection{The contribution of the four-point functions of primary operators.}

For a general CFT the four-point function has the form
\be\fl
\langle \phi_{k_1}(z_1) \phi_{k_2}(z_2) \phi_{k_3}(z_3) \phi_{k_4}(z_4)\rangle_{\bf C}=
\prod_{1\leq q<p\leq4} (z_p-z_q)^{\Delta_{pq}}(\zb_p-\zb_q)^{\Db_{pq}} G(x_z,\bar x_z)\,,
\ee
with $\Delta_{pq}\equiv \sum_{\ell=1}^4 \Delta_\ell/3-\Delta_p-\Delta_q$ and similarly for $\Db_{pq}$.
$x_z$ stands for the four point ratio of the points $z_p$.
The universal function $G(x_z,\bar x_z)$ should be calculated on a case by case basis because it depends on the CFT and on the 
specific operators considered \cite{bpz-84,book}. 
The function $G(x_z,\bar x_z)$ can be rather complicated for a general CFT, but in the two cases we will consider in the following and for 
the operators of lower dimension, it assumes a rather simple form that allows to write down $d_{\{k_j\}}$ explicitly. 

The four spin correlation function in the Ising model can be written as \cite{bpz-84,book}
\be\fl
\langle \sigma(z_1) \sigma(z_2) \sigma(z_3) \sigma(z_4)\rangle_{\bf C}^2=\frac14 \left|
\frac{z_{13} z_{24}}{z_{14}z_{23}z_{12}z_{34}}\right|^{1/2}\left[1+\left| \frac{z_{12}z_{34}}{z_{13}z_{24}}\right|
+\left|\frac{z_{14}z_{23}}{z_{13}z_{24}}\right|\right]\,,
\label{4ptIs}
\ee
that gives the lowest power in $x$ coming from a four-point function in the short length expansion. 
For the free  compactified boson, the lowest contribution is given by the correlation function of four vertex operator 
two with charge $\sqrt\mu$ ($V_\mu(z)=e^{i\sqrt\mu \varphi(z)}$) and two $-\sqrt\mu$ ($V_{-\mu}(z)=e^{-i\sqrt\mu \varphi(z)}$).
The value of $\mu=\a=\min[\eta,1/\eta]$ is directly related to the compactification radius.
The three relevant correlations are
\bea
\langle V_\mu(z_1) V_\mu(z_2) V_{-\mu}(z_3) V_{-\mu}(z_4)\rangle_{\bf C}&=& \left|
\frac{z_{12} z_{34}}{z_{14}z_{23}z_{14}z_{23}}\right|^{2\mu}\,, 
\nonumber\\
\langle V_\mu(z_1) V_{-\mu}(z_2) V_{\mu}(z_3) V_{-\mu}(z_4)\rangle_{\bf C}&=& \left|
\frac{z_{13} z_{24}}{z_{14}z_{23}z_{12}z_{34}}\right|^{2\mu}\,, 
\nonumber\\
\langle V_\mu(z_1) V_{-\mu}(z_2) V_{-\mu}(z_3) V_{\mu}(z_4)\rangle_{\bf C}&=& \left|
\frac{z_{23} z_{14}}{z_{24}z_{13}z_{12}z_{34}}\right|^{2\mu}\,. 
\eea
The correlators with $\mu\to-\mu$ should be also included and these give identical contributions.
The correlations for the Ising model and for the free boson can be treated on the same foot since the formers come from
the bosonization of the latter with $\mu=1/4$.

It is now easy to obtain the factors $d_{\mu_1 \mu_2 \mu_3 \mu_4}^{(j_1j_2j_3j_4)}$ with $\mu_p=\pm \mu$ as for example
\bea\fl
d_{\mu, \mu, -\mu, -\mu}^{(j_1 j_2 j_3 j_4)} &=&
n^{-4\mu}
\langle V_{\mu}(e^{\frac{2\pi i j_1}{n}}) V_{\mu}(e^{\frac{2\pi i j_2}{n}})
V_{-\mu}(e^{\frac{2\pi i j_3}{n}}) V_{-\mu}(e^{\frac{2\pi i j_4}{n}})\rangle_{\bf{C}}\nonumber\\ \fl
&=&(2n)^{-4\mu}
\left|\frac{\sin(\pi j_{21}/n) \sin(\pi j_{43}/n)}{\sin(\pi j_{42}/n) \sin(\pi j_{31}/n) \sin(\pi j_{41}/n) \sin(\pi j_{32}/n)}\right|^{2\mu}
\,,
\label{d4}
\eea
and the other two just by permuting properly $j_p$. We used that all these operators have zero conformal spin.

\section{The Ising model}
\label{SecIM}

The partition function of the Ising model on a generic Riemann surface of genus $g$ with {\it period matrix} $\tau$ has been 
obtained in Ref.  \cite{Dijkgraaf:1987vp} (see also \cite{orb2}) and reads
\be
Z= (Z_0^{\rm qu})^{1/2} 2^{-g} \sum_{\bm{\varepsilon}, \bm{\delta}} 
\left|\Theta\bigg[\begin{array}{c} \bm{\varepsilon} \\ \bm{\delta} \end{array}\bigg] ({\bf 0}|\tau)\right|\,,
\label{dvv}
\ee
where $Z_0^{\rm qu}$ is the quantum part of the partition function for a free boson 
that does not depend on the details of the Riemann surface  (and on the target space as well) but only on its topology. 
$Z_0^{\rm qu}$ has been calculated by Dixon et al. \cite{Dixon:1986qv} (see also \cite{z-87}).  
In Eq. (\ref{dvv}), the sum is intended over all the $2^g$ vectors $\bm{\varepsilon}$ and $\bm{\delta}$ with entries $0$ and $1/2$. 
We mention that this result for the Ising model is rather special, indeed usually it is {\it not}  possible to give a closed 
formula for a general Riemann surface with arbitrary period matrix $\tau$ for all CFTs.
For example, also for the compactified  boson, such a formula does not exist (in \cite{cct-09}, the property that the matrix $\tau$ is purely 
imaginary has been exploited to obtain Eq. (\ref{Fnv})). 

Combining Eq. (\ref{dvv}) with our results in Ref. \cite{cct-09},   it is easy to derive 
the Ising partition function on the $n$-sheeted Riemann surface ${\cal R}_{n,2}$. 
In fact, a by-product of  the calculation for the free boson  \cite{cct-09} is that the $(n-1)\times (n-1)$ period matrix is given  
by Eq. (\ref{Gammadef}).
Although derived for a free boson, the period matrix is a pure geometrical object and it is only related to the structure of the world-sheet 
${\cal R}_{n,2}$ and so it is the same for any theory.

The function ${\cal F}_n(x)$ for the Ising model is proportional to Eq. (\ref{dvv}) with $\tau$ equal to the matrix $\Gamma$ in Eq. 
(\ref{Gammadef}), and 
the proportionality constant is fixed by requiring ${\cal F}_n(0)=1$. In this way, we simply obtain Eq. (\ref{Fn ising}). 
We mention that for known formulas analogous to Eq. (\ref{dvv}) and valid for other theories, 
${\cal F}_n(x)$ can be easily obtained in the same way. 

\begin{figure}
\includegraphics[width=0.8\textwidth]{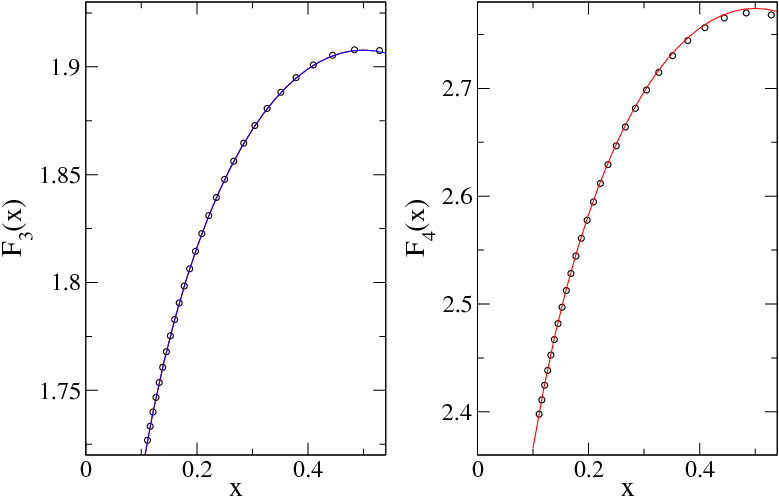}
\caption{${\cal F}_{3,4}(x)$: The extrapolated numerical data of Ref. \cite{fc-10} are compared with our prediction. Only $x\leq 0.5$ are reported because of the symmetry $x\to1-x$. }
\label{fig-FnIS}
\end{figure}

We performed several checks for the correctness of this equation. 
For $n=2$ it reduces to 
\be\fl
{\cal F}_2(x) = \frac1{\sqrt{2}}\Bigg[
\left[\frac{(1 + \sqrt{x}) (1 + \sqrt{1 - x})}2\right]^{1/2} 
+ x^{1/4} + [(1 - x) x]^{1/4} + (1 - x)^{1/4} \Bigg]^{\frac12},
\label{F2Is}
\ee
obtained (and checked against numerical calculations) in Ref. \cite{atc-09}.
For $n=3,4$ numerical estimates of these functions have been obtained from the exact diagonalization of the Ising chain in \cite{fc-10}.
The comparison of our prediction with the numerical data for $n=3$ and $n=4$ is reported in Fig. \ref{fig-FnIS}. 
The agreement is excellent, taking also into account that the
extrapolation to $\ell\to\infty$ in Ref. \cite{fc-10} has been obtained without knowing the asymptotic result.
 
\subsection{Invariance under $x \rightarrow 1-x$}

When $x \rightarrow 1-x$, ${\cal F}_n(1-x)$ is given by Eq. (\ref{Fn ising}) where $\Gamma$ is replaced by  $\widetilde{\Gamma}$ given  
by Eq. (\ref{Gammadef}) where $x$ is replaced by $1-x$ (i.e.~$\beta_y\to1/\beta_y$). 
We denote by $\widetilde{\cal F}_n(x)={\cal F}_n(1-x)$ the expression 
obtained by using $\widetilde{\Gamma}$ instead of $\Gamma$ in (\ref{Fn ising}).
In order to show the invariance of (\ref{Fn ising}) under $x \rightarrow 1-x$, we observe that in general we have
\begin{equation}
\label{theta transform}
\Theta
\bigg[\begin{array}{c} {\bf a} \\ {\bf b}  \end{array}\bigg] ({\bf z} |T)
\,=\,
\frac{e^{2\pi i\, {\bf a}^{{\rm t}}\cdot ({\bf z+b})}}{\sqrt{{\rm det}(-i T)}}\;
\Theta
\bigg[\begin{array}{c}{\bf  z+b} \\{\bf -a}  \end{array}\bigg] ({\bf 0} |-T^{-1})\,,
\end{equation}
for all complex vectors  ${\bf a\,, b}$ and ${\bf z}$. This can  be shown by a straightforward generalization of the first part of the appendix B of
Ref. \cite{cct-09}.
The relation (\ref{theta transform}) implies that
\begin{equation}
\label{theta transform2}
\frac{1}{\Theta({\bf 0}|T)} \Big| \Theta\bigg[\begin{array}{c} {\bf a} \\ {\bf b}  \end{array}\bigg] ({\bf 0}|T) \Big|
=
\frac{1}{\Theta({\bf 0}|-T^{-1})}\, \Big| \Theta\bigg[\begin{array}{c} {\bf b} \\ -{\bf a}  \end{array}\bigg] ({\bf 0}|-T^{-1}) \Big|\,.
\end{equation}
Applying this relation to $T = \widetilde{\Gamma}=-A\Gamma^{-1} A$ (the transformation matrix $A$ is defined in 
App. B of Ref. \cite{cct-09}), we find
\begin{equation}
\widetilde{\cal F}_n(x) = 
\frac{1}{2^{n-1}\Theta({\bf 0}|A^{-1}\Gamma A)} \sum_{\bm{\varepsilon,\delta}}
\Big| \Theta\bigg[\begin{array}{c} \bm{\delta} \\  -\bm{\varepsilon}   \end{array}\bigg] ({\bf 0}|A^{-1}\Gamma A) \Big|\,.
\end{equation}
In Ref. \cite{cct-09} we have used that $\Theta({\bf 0}|A^{-1}\Gamma A)=\Theta({\bf 0}|\Gamma)$ without proof (see Eq. (B.5) of \cite{cct-09}). 
Here we argue that a generalized version of this equation is  
\begin{equation}
\sum_{\bm {\varepsilon,\delta}}
\Big| \Theta\bigg[\begin{array}{c}  {\bm \delta} \\  -{\bm \varepsilon}   \end{array}\bigg] ({\bf 0}|A^{-1}\Gamma A) \Big|
=
\sum_{\bm{\varepsilon,\delta}}
\Big| \Theta\bigg[\begin{array}{c} {\bm \varepsilon}  \\ {\bm \delta}    \end{array}\bigg] ({\bf 0}|\Gamma) \Big|\,,
\end{equation}
which is numerically true and we give without proof. From this relation, we find that (\ref{Fn ising}) is invariant under 
$x \rightarrow 1-x$, namely ${\cal F}_n(1-x)=\widetilde{\cal F}_n(x)={\cal F}_n(x)$.

\section{The small $x$ expansion}
\label{Secsx}

As we showed above, for small $x$, the scaling function ${\cal F}_n(x)$ displays a systematic expansion revealing the full 
operator content of the theory in an explicit manner. 
Here we consider the first two non vanishing contributions from the expansion of  ${\cal F}_n(x)$ 
for the free boson and the Ising model. 
This expansion has the form reported in Eq. (\ref{Fexpintro})
that gives the first terms in the asymptotic expansion as long as $2\alpha<1$, because $O(x)$ terms exist as well (the latter are
actually trivial as we shall see and do not correspond to any operator).
We recall that $\a$ is the double of the leading scaling operator in the theory (i.e.~for the boson $\a=\min[\eta,1/\eta]$ and for Ising 
$\a=1/4$). 
The coefficients $s_{2,4}(n)$ are explicitly calculated in the following. 
We mention that in both models the contribution of non vanishing three-point functions are subleading, as clear from the 
conformal spectra of the two theories.

Notice that, since the exponents in this expansion do not depend on $n$, the structure of the power series passes over to the 
analytic continuation and so the scaling function of the von Neumann entanglement entropy is
\be\fl
{\cal F}_{VN}(x)=\frac{\partial}{\partial n} {\cal F}_n (x)\Big|_{n=1}=\left(\frac{x}4\right)^{\a} s'_2(1)
+\left(\frac{x}4\right)^{2\a}  s'_4(1)
+\cdots\,,
\ee
where we used ${\cal F}_1(x)=1$ (i.e.~$\Tr\rho_A=1$). 
The calculation of the analytic continuation of the amplitude $s_m(n)$ and their derivative at $n=1$ 
is rather complicated and will be considered in Sec.~\ref{ancont}.

\subsection{First order for the compactified boson}

The calculation of the leading coefficient $s_2(n)$ for a free compactified boson 
has been already reported in Ref. \cite{cct-09}. This is repeated here for completeness and to fix the notation for the 
Ising model and for calculating the second non vanishing order.

The first ingredient we need is the expansion of $\beta_{k/n}$ for $x \to 0$ 
$$
 \beta_{\frac{k}{n}} =
-\frac{\sin\left(\pi\frac{k}{n}\right)}{\pi}
\left(\log x+ 
f_{\frac{k}{n}}
+\sum_{l=1}^{\infty} p_l\left(\frac{k}{n}\right)x^l\right),
\hspace{.2cm}
f_{\frac{k}{n}} \equiv 
2\gamma_E +\psi\left(\frac{k}{n}\right)+ \psi\left(1-\frac{k}{n}\right),
$$
where $\gamma_E$ is the Euler gamma, $\psi(z)\equiv\Gamma'(z)/\Gamma(z)$ is 
the polygamma function 
and $p_l(z)$ is a 
polynomial of degree $2l$, whose explicit expression is not needed because it contributes at order $o(x)$ that we do not consider here. 
Plugging this expansion in the Riemann-Siegel theta function we obtain
\bea
\fl \Theta({\bf 0}|\eta \Gamma) =
1+
\sum_{{\bf m \,\in\,  Z}^{n-1}\setminus {\bf \{0\}}}
x^{\eta a_{\bf m}} e^{\eta b_{\bf m}} \big[1+O(x)\big]\,,
\label{abdefbos}
\eea
where the two vector-dependent coefficients $a_{\bf m}$ and $b_{\bf m}$ are
\begin{eqnarray}
\label{alphadef}
a_{\bf m} &\equiv & 
\frac{2}{n} \sum_{k=1}^{n-1} \sin^2\left(\pi\frac{k}{n}\right) \,{\bf m}^{{\rm t}}\cdot   C_{k/n} \cdot {\bf m}\,,\\
\label{bdef}
b_{\bf m} &\equiv &
\frac{2}{n} \sum_{k=1}^{n-1}  \sin^2\left(\pi\frac{k}{n}\right) f_{k/n}\, {\bf m}^{{\rm t}}\cdot   C_{k/n} \cdot {\bf m}\,,
\end{eqnarray}
and the matrix $C_{k/n}$ has elements \cite{cct-09}
\begin{equation}
\left( C_{\frac{k}{n}} \right)_{rs}\,\equiv\, 
 \cos\left[2\pi\frac{k}{n}(r-s)\right]\,.
\label{CCdef}
\end{equation}
The leading term in the small $x$ expansion is provided by those vectors 
${\bf m} \in  {\bf Z}^{n-1} \setminus {\bf \{0\}}$
which minimize $a_{\bf m}$. When evaluated on the vectors ${\bf m}$, it  is easy to show that 
\begin{equation}
\fl a_{\bf m}= 
\sum_{j=1}^{n-1} m_j^2 -\sum_{j=1}^{n-2} m_j \,m_{j+1}=
\frac{m_1^2+m_{n-1}^2}2+\frac12 \sum_{j=1}^{n-2}(m_j-m_{j+1})^2
\,,
\label{am2}
\end{equation}
where $m_j$ indicates the $j$ component of the vector ${\bf m}$.
When minimizing $a_{\bf m}$, the second term in the rhs of the last expression tends to make all the components of ${\bf m}$ equal 
and only ``domain walls'' give a non vanishing contribution, proportional to the square of the difference of the two neighbor  
components  at the wall. 
The first term is like a ``boundary term'' and tend to make the absolute value of of the boundaries $m_1$ and $m_{n-1}$ as small
as possible. 
Within these ingredients, it is obvious that (among the non zero vectors) $a_{\bf m}$ is minimized by all vectors of the form 
\begin{equation}
\label{mvectors order1}
{\bf m}^{{\rm t}}_{\pm} \equiv   \big( \underbrace{0, \dots, 0 }_{j_1},  \underbrace{\pm 1, \dots, \pm 1 }_{j_2-j_1}, 0, \dots ,0 \big)\,,
\end{equation}
with $0\leq j_1<j_2\leq n-1$ that give all $a_{\bf m}=1$.
In order to calculate $b_{\bf m}$, we use
\begin{equation}\fl
s(j_2-j_1)\equiv {\bf m}^{{\rm t}}_{\pm} \cdot  C_{k/n} \cdot  {\bf m}_{\pm}
= \sum_{p,q=1}^{j_2-j_1} \left(C_{\frac{k}{n}} \right)_{pq}
=\left(\frac{\sin\left(\pi\frac{k}{n} (j_2-j_1)\right)}{\sin\left(\pi\frac{k}{n}\right)}\right)^2\,,
\label{sl1}
\end{equation}
and the  integral representation of the polygamma function
\begin{equation}
\psi(y) +\gamma_E= \int_0^\infty\frac{e^{-t}-e^{-yt}}{1-e^{-t}}\,dt\,.
\label{intrep}
\end{equation}
Exchanging the order of sum and integral and then performing the latter, 
after simple algebra we find
\begin{equation}
b_{{\bf m}_{\pm}}=-2\log\left[2n \sin\left(\pi\frac{j_2-j_1}{n}\right)\right]\,.
\label{bm}
\end{equation}
We now can use the translational invariance of $s(j_2-j_1)$ to fix $j_1=0$ and multiply by the number of possible $j_1$ at fixed
$j_2-j_1$, i.e.~$(n-j_2)$. There is another factor $2$ to take into account  the two possible signs in the vectors ${\bf m}_{\pm}$. 

Now, assuming $\eta \neq 1$, in the ratio of Riemann-Siegel theta functions occurring in 
$\mathcal{F}_n(x)$ the leading term is given by the minimum between $\eta$ and $1/\eta$, that 
we indicate as $\alpha={\rm min}(\eta,1/\eta)$. 
Therefore we get the small $x$ behavior of the scaling function (renaming $j_2=j$) 
\bea
\fl\mathcal{F}_n(x)& =&1+
x^\alpha
\sum_{j=1}^{n-1}
\frac{2(n-j)}{\left[ 2n \sin\left(\pi\frac{j}{n}\right) \right]^{2\alpha}} + \dots=
1+ 2\left(\frac{x}{4n^2}\right)^\alpha
\sum_{j=1}^{n-1}
\frac{j}{\left[ \sin\left(\pi\frac{j}{n}\right) \right]^{2\alpha}}+ \dots \nonumber\\ \fl &=&
1+ n\left(\frac{x}{4n^2}\right)^\alpha
\sum_{j=1}^{n-1}
\frac1{\left[ \sin\left(\pi\frac{j}{n}\right) \right]^{2\alpha}}+ \dots
\,, \label{Fsmallx}
\eea
where the dots denote higher order terms in $x$. 
We stress  that this expansion is valid only for $\eta\neq1$, in fact for 
$\eta=1$ the denominator in ${\cal F}_n(x)$ (that is of order $O(x)$) cancels exactly 
the numerator. 

This first order term in Eq. (\ref{Fsmallx}) coincides exactly with the prediction from short length expansion 
(\ref{s2sle}) with $x_k=\a/2$ corresponding to the scaling dimension of the vertex operators $V_{\pm \a}(z)$.
The additional factor $2$ comes from the fact that both correlations $\langle V_\a V_{-\a}\rangle$ and 
$\langle V_{-\a} V_{\a}\rangle$ must be considered.

\subsection{Digression on the denominator}

The previous analysis also provides the small $x$ expansion of the denominator in ${\cal F}_{n}(x)$ corresponding to $\a=1$.
We have 
\be
\Theta({\bf 0}|\Gamma)=1+  n\left(\frac{x}{4n^2}\right)
\sum_{j=1}^{n-1}
\frac1{\left[ \sin\left(\pi\frac{j}{n}\right) \right]^{2}}+ \cdots\,.
\ee
As showed in Ref. \cite{cct-09} we have
\be
 \sum_{l=1}^{n-1}
\frac{1}{\left[\sin\left(\pi{l}/{n}\right) \right]^{2 }} =
\frac{1}3 (n^2-1)\,,
\ee
so that the denominator is 
\be
[\Theta({\bf 0}|\Gamma)]^2=1+ x \,\frac16\left(n-\frac1n\right)+O(x^2)\,,
\label{denexp}
\ee
that gives an $O(x)$ contribution to the scaling function ${\cal F}_n(x)$.
However there is no operator in the spectrum of the free boson whose correlation function can contribute  at order $O(x)$. 
This fact could sound as a breakdown of the short-length expansion, but it is not. 
Indeed, it predicts the asymptotic expansion of the full $\Tr\rho_A^n$ and in general any term 
in this expansion is also reproduced in ${\cal F}_n(x)$. 
However, in the case of integer powers, we should properly take into account the prefactor of ${\cal F}_n(x)$ in $\Tr\rho_A^n$
as in Eq. (\ref{Fn}).
According to Eq. (\ref{pref}), this is (for $c=1$)
\be\fl
\left[\frac1{\ell_1\ell_2(1-x)}\right]^{1/6 (n-1/n)}=
\frac1{(\ell_1\ell_2)^{1/6 (n-1/n)}}\left(1+x \frac16(n-1/n)+O(x^2)\right)\,,
\ee
that exactly cancels the $O(x)$ contribution of the denominator in Eq. (\ref{denexp}).
Thus there is no $O(x)$ contribution in $\Tr\rho_A^n$ in agreement with the short-length expansion.

\subsection{First order in $x$ for the Ising model}

\label{first order ising}

We can now move to the calculation of the first non vanishing contribution to ${\cal F}_n(x)$ for the Ising model.
In order to study the behavior of Eq. (\ref{Fn ising}) when $x \to 0$ we isolate the contribution of the terms with 
${\bm \varepsilon}={\bf 0}$
\begin{eqnarray}\fl
{\cal F}_n(x) &=&
\frac{1}{2^{n-1}\Theta({\bf 0}|\Gamma)} \bigg(
\sum_{\bm \delta}
\Big| \Theta\bigg[\begin{array}{c} {\bf 0} \\ {\bm \delta}  \end{array}\bigg] ({\bf 0}|\Gamma) \Big|
+
\sum_{{\bm \varepsilon}\neq {\bf 0}} \sum_{\bm \delta}
\Big| \Theta\bigg[\begin{array}{c} {\bm \varepsilon} \\ {\bm \delta}  \end{array}\bigg] ({\bf 0}|\Gamma) \Big|
\bigg)\\ \fl
\label{Fn x0 step1}
&=&
\frac{1+O(x)}{2^{n-1}} \bigg(
\sum_{\bm \delta} \big[1+O(x)\big]
+
\sum_{{\bm \varepsilon}\neq {\bf 0}} \sum_{\bm \delta}
\Big| \Theta\bigg[\begin{array}{c} {\bm \varepsilon} \\ {\bm \delta}  \end{array}\bigg] ({\bf 0}|\Gamma) \Big|
\bigg)\,.
\end{eqnarray}
By using the definition of $a_{\bf m}$ and $b_{\bf m}$ in the previous subsection, we find for ${\bm \varepsilon}\neq {\bf 0}$
\begin{equation}
\Theta\bigg[\begin{array}{c} \bm\varepsilon \\ \bm\delta  \end{array}\bigg] ({\bf 0}|\Gamma)
=
\sum_{{\bf m} \in  \mathbf{Z}^{n-1}} e^{2\pi i ({\bf m}+\bm\varepsilon)^{{\rm t}}\cdot \,\bm\delta}
x^{a_{\bf m+\bm\varepsilon}} e^{b_{\bf m+\bm\varepsilon}} \big[1+O(x)\big]\,.
\end{equation}
Again the leading contribution is given by all the vectors ${\bf m}$ that  minimize $a_{\bf m}$.
By simple inspection, the minimum value of $a_{\bf m+\bm \varepsilon}$ is $1/4$ and for any $\bm\varepsilon$ of the form 
\begin{equation}
\bm\varepsilon^{{\rm t}} =  \big( \underbrace{0, \dots, 0 }_{j_1},  \underbrace{1/2, \dots, 1/2 }_{j_2-j_1}, 0, \dots ,0 \big)\,,
\end{equation}
the value $a_{\bf m+\bm \varepsilon} =1/4$ is obtained for the two vectors
\begin{eqnarray}\fl
{\bf m}^{{\rm t}} \equiv 
 \big( \underbrace{0, \dots, 0 }_{j_1},  \underbrace{-1, \dots, -1 }_{j_2-j_1}, 0, \dots ,0 \big)
 &\Longrightarrow & 
 ({\bf m}+\bm \varepsilon)^{{\rm t}} = \frac{1}{2}  {\bf m}^{{\rm t}}_{-}\,,
 \\ \fl
{\bf m}^{{\rm t}} \equiv 
 \big( 0, \dots ,0 \big)\hspace{3.5cm}
 &\Longrightarrow & 
 ({\bf m}+\bm \varepsilon)^{{\rm t}} = \frac{1}{2}  {\bf m}^{{\rm t}}_{+}\,,
\end{eqnarray}
where ${\bf m}_{\pm}$ are the vectors employed for the free boson in Eq. (\ref{mvectors order1}).
It is therefore clear that we can exploit the results of the free boson to obtain the expansion for Ising very easily. 
In particular, $a_{\bf m+\bm\varepsilon} $ and $b_{\bf m+\bm\varepsilon} $ are $1/4$ of the corresponding free boson quantities.
Thus from Eq. (\ref{Fn x0 step1}) we have
\begin{eqnarray}\fl
{\cal F}_n(x) &=&
\frac{1}{2^{n-1}} \bigg(2^{n-1}
+
\sum_{\bm\varepsilon\neq \bf0} \sum_{\bm\delta}
\Big| \Theta\bigg[\begin{array}{c} \bm\varepsilon \\ \bm\delta  \end{array}\bigg] ({\bf 0}|\Gamma) \Big|
\bigg) + \dots \\ \fl
&=&
1+\frac{1}{2^{n-1}}  \sum_{\bm \varepsilon\neq \bf0} \sum_{\bm\delta}\;
\Big|\sum_{{\bf m \in Z}^{n-1}} e^{2\pi i ({\bf m}+\bm\varepsilon)^{{\rm t}}\cdot \,\bm\delta}
x^{a_{\bf m+\bm\varepsilon}} e^{b_{\bf m+\bm\varepsilon}}  \Big|
+\dots \label{deltaeff} \\ \fl
&=&
1+ \frac{x^{1/4}}{2^{n-1}}  \sum_{0\leq j_1<j_2\leq n-1}
e^{b_{({\bf m}+\bm\varepsilon)_{{\pm}}}}
\sum_{\bm\delta}
\Big|e^{\pi i {\bf m}_{+}^{{\rm t}}\cdot \,\bm\delta} + e^{\pi i {\bf m}_{-}^{{\rm t}}\cdot \,\bm\delta}  \Big|
+\dots \\ \fl
\label{Fn x0 step2}
&=&
1+ \frac{x^{1/4}}{2^{n-1}}  \sum_{j = 1}^{n-1} 
 (n-j) e^{b_{({\bf m}+\bm\varepsilon)_{\pm}}}
\sum_{\bm\delta}
\Big|e^{\pi i {\bf m}_{+}^{{\rm t}}\cdot \,\bm\delta} + e^{-\pi i {\bf m}_{+}^{{\rm t}}\cdot \,\bm\delta}  \Big|
+\dots \\ \fl
\label{Fn x0 step3}
&=&
1+ \frac{x^{1/4}}{2^{n-1}}  \sum_{j = 1}^{n-1} 
\frac{(n-j)}{\big[2n \sin(\pi j/n)\big]^{2\frac{1}{4}}} 
\sum_{\bm \delta}
\Big|e^{\pi i {\bf m}_{+}^{{\rm t}}\cdot \,\bm\delta} + e^{-\pi i {\bf m}_{+}^{{\rm t}}\cdot \,\bm\delta}  \Big|
+\cdots\,. 
\end{eqnarray}
The various steps are very easy, but it is pedagogical to summarize the properties we employed.  
We used that $b_{({\bf m}+\bm\varepsilon)_\pm}$ is real and independent of $\bm\delta$, $\pm$ and $j_1$. 
In (\ref{Fn x0 step2}) we have used the translational invariance to fix $j_1=0$ and renamed $j_2=j$. 
In (\ref{Fn x0 step3}) we have employed the free-boson result  exposing the role of the factor $1/4$ which is relevant in this case.
The next step is to  observe that ${\bf m}_{+}^{{\rm t}}\cdot \bm\delta= q/2$, where $q$ is the number of $1/2$'s occurring 
in the first $j$ coordinates of $\bm\delta$. 
The remaining $n-1-j$ coordinates of $\bm\delta$ do not change the scalar product 
and this freedom provides a factor $2^{n-1-j} $. Thus we find
\begin{equation} \fl
\sum_{\bm \delta}
\Big|e^{\pi i {\bf m}_{+}^{{\rm t}}\cdot \bm\delta} + e^{-\pi i {\bf m}_{+}^{{\rm t}}\cdot \bm\delta}  \Big|
= 2^{n-1-j} \sum_{q=0}^j 
{j \choose q} \left| e^{\pi i q/2} + e^{-\pi i q/2}  \right|
= 2^{n-1}\,.
\end{equation}
The final result is
\begin{eqnarray}\fl
{\cal F}_n(x) = 1+ x^{\frac14}  \sum_{j = 1}^{n-1} 
\frac{(n-j)}{\big[2n \sin(\pi j/n)\big]^{\frac{1}{2}}} +\cdots =
1+ 
\frac{n}2 \left[\frac{x}{4n^2}\right]^{\frac14}
 \sum_{j = 1}^{n-1} 
\frac{1}{\sin^{\frac{1}{2}}[\pi j/n]} 
+\cdots.
\label{F1 ising final}
\end{eqnarray}
Notice that this equation has the same form as the one for free boson Eq. (\ref{Fsmallx}) for $\a=1/4$, 
with an additional factor $1/2$ in front.
This is exactly what predicted by short length expansion (cf. Eq. (\ref{s2sle})) from the two-point spin correlation function.

As for the free boson, the $O(x)$ contribution from the denominator of ${\cal F}_n(x)$ cancels with the $O(x)$ contribution of the 
prefactor in Eq. (\ref{Fn}), in agreement with the short length expansion, since also for the Ising model no correlation 
function can contribute at $O(x)$. 

\subsection{Second order for the compactified boson}
\label{second order small x boson}

To obtain the second order in the small $x$ expansion, we need to know the vectors ${\bf m}$ in Eq. (\ref{abdefbos}) that give 
the second minimal value of $a_{\bf m}$. 
Since $a_{\bf m}$ can assume only integer values (cf. Eq. (\ref{am2})), we need to identify all vectors giving $a_{\bf m}=2$. 
From the rhs of Eq. (\ref{am2}), we can have $a_{\bf m}=2$ only by introducing another two domain-walls of height one 
in the string of $\pm1$ in ${\bf m}_{\pm}$ in Eq. (\ref{mvectors order1}). 
Thus the vector ${\bf m}$ giving $a_{\bf m}=2$ can only be of the forms
\begin{eqnarray} \fl
\label{mvecdef}
 {\bf m}_0^{{\rm t}} &\equiv &
 \big( \underbrace{0, \dots, 0 }_{j_1}, {\underbrace{\pm 1, \dots, \pm 1}_{j_2-j_1}, 
  \underbrace{0, \dots, 0}_{j_3-j_2}, \underbrace{\pm1, \dots, \pm1}_{j_4-j_3}}, 0, \dots ,0\big)
  \,,
 \\ \fl
 {\bf m}_1^{{\rm t}} &\equiv &
 \big( \underbrace{0, \dots, 0 }_{j_1}, {\underbrace{\pm 1, \dots, \pm 1}_{j_2-j_1}, 
  \underbrace{0, \dots, 0}_{j_3-j_2}, \underbrace{\mp1, \dots, \mp1}_{j_4-j_3}}, 0, \dots ,0\big)
  \,,\nonumber
\\ \fl
{\bf m}_2^{{\rm t}}  &\equiv &
 \big( \underbrace{0, \dots, 0 }_{j_1}, \underbrace{\pm 1, \dots, \pm 1}_{j_2-j_1}, 
 \underbrace{ \pm 2, \dots, \pm 2}_{j_3-j_2} , \underbrace{\pm 1, \dots, \pm 1}_{j_4-j_3}, 0, \dots ,0\big)\,,\nonumber
\end{eqnarray}
where $0\leq j_1<j_2<j_3<j_4\leq n-1$. 
For each ${\bf m}_r$ we have two types of vectors corresponding to the sign of the first non-zero element (when needed we will 
specify the vector with an additional $\pm$ subscript).
We remark that these vectors  exist only for $n \geq 4$. 
{\it For $n=2$ and $n=3$ the minimum value of $a_{\bf m}$ greater than 1 is $a_{\bf m}=4$ and $a_{\bf m}=3$ respectively, as 
predicted by short-length expansion.}

In order to  get $b_{\bf m}$ for the vectors ${\bf m}_r$  above, the first step  
is to compute ${\bf m}_r^{{\rm t}} \cdot C_{k/n} \cdot {\bf m}_r$. 
We introduce the sum of the elements of the matrix $C_{k/n}$ enclosed in a rectangle made by 
$\lambda_r$ rows and $\lambda_c$ columns starting at the $(p+1)$-th row and at the $(q+1)$-th column. It is given by
\begin{eqnarray} \fl
s(p,\lambda_r, q,\lambda_c)  &\equiv&
\sum_{i=p+1}^{p+\lambda_r} \, \sum_{j=q+1}^{q+\lambda_c}  (C_{k/n})_{ij}\\ \fl
&=&\frac{\sin (\pi(k/n)\lambda_r) \sin (\pi(k/n)\lambda_c)}{\sin (\pi(k/n))^2}\, 
\cos \bigg(\pi\frac{k}{n}\big(2[p-q]+\lambda_r-\lambda_c\big) \bigg)\\
 \fl
& =  &
\frac{1}{2\sin^2 (\pi k/n)}
\left\{ \sin^2 \bigg[\frac{\pi k}{n}(p-q+\lambda_r) \bigg]
+\sin^2 \bigg[\frac{\pi k}{n}(p-q-\lambda_c) \bigg] \right. \nonumber  \\ \fl
& & \hspace{1.3cm} \left.-\;\sin^2 \bigg[\frac{\pi k}{n}(p-q+\lambda_r-\lambda_c) \bigg]
-\sin^2 \bigg[\frac{\pi k}{n}(p-q) \bigg]
 \right\}\,.\label{sum Ckn step2}
\end{eqnarray}
In the special case of a square (i.e.~$\lambda_r=\lambda_c\equiv \lambda$) on the diagonal (i.e.~$p=q$) it reduces to
Eq. (\ref{sl1}), i.e.~$s(p,\lambda,p,\lambda)=s(\lambda)$.
Then, we have
\begin{eqnarray} \fl
\label{mjCknmj}
{\bf m}_0^{{\rm t}} \cdot C_{k/n} \cdot {\bf m}_0 
&=&
s(j_{21})+s(j_{43})+2  s(j_1,j_{21},j_3,j_{43})\,,  \\ \fl
{\bf m}_1^{{\rm t}} \cdot C_{k/n} \cdot {\bf m}_1 
&=&
s(j_{21})+s(j_{43})-2  s(j_1,j_{21},j_3,j_{43})\,,\nonumber \\ \fl
\label{m2 Ckn m2}
{\bf m}_2^{{\rm t}} \cdot C_{k/n} \cdot {\bf m}_2 &=&
s(j_{21})+ s(j_{43}) + 2 s(j_3,j_{43} , j_1,  j_{21}) \nonumber\\ \fl
& &+ 4 ( s(j_{32})+s(j_2 ,j_{32} , j_3,j_{43})+  s(j_1,j_{21} , j_2, j_{32}))
\,, \nonumber 
\end{eqnarray}
where we introduced $j_{kl}=j_k-j_l$.
 
 At this point we have all the ingredients to calculate $b_{{\bf m}_r}$  from their definition (\ref{bdef}) and from 
 Eq. (\ref{intrep}) to deal with the polygamma function. 
 It is a tedious, but straightforward calculation to obtain these coefficients.
 After long algebra, we finally have
 \be
 b_{{\bf m}_{r}}
= \log  \left(\frac{Q_{r}}{4n^2}\right)^2\,,
\hspace{1cm} {\rm for}\; r=0,1,2\,,
\label{bmj}
 \ee
where we defined 
\begin{eqnarray}
\label{Q1p def}
Q_{0} &\equiv& 
\frac{\sin(\pi j_{42}/n) \sin(\pi j_{31}/n)}{\sin(\pi j_{21}/n) \sin(\pi j_{43}/n) \sin(\pi j_{41}/n) \sin(\pi j_{32}/n)}\,,\\ 
\label{Q1m def}
Q_{1} &\equiv& 
\frac{\sin(\pi j_{41}/n) \sin(\pi j_{32}/n)}{\sin(\pi j_{21}/n) \sin(\pi j_{43}/n) \sin(\pi j_{42}/n) \sin(\pi j_{31}/n)}\,,\\ 
\label{Q2 def}
Q_{2} &\equiv& 
\frac{\sin(\pi j_{21}/n) \sin(\pi j_{43}/n)}{\sin(\pi j_{42}/n) \sin(\pi j_{31}/n) \sin(\pi j_{41}/n) \sin(\pi j_{32}/n)}\,.
\end{eqnarray}
We have $Q_r>0$ for any $r$.
Notice that $Q_{0,1,2}$ are invariant under translation (i.e.~depend only on the differences of the $j$'s) although 
Eqs. (\ref{mjCknmj}) are not.
There are few other important symmetries of these factors.
Each $Q_r$ is invariant under $1 \leftrightarrow 2$ and $3 \leftrightarrow 4$. 
Under the cyclic permutation $\{1,2,3,4\}\leftrightarrow \{4,1,2,3\}$, we have $Q_0\to -Q_0$ and $Q_{1,2}\to -Q_{2,1}$,
but the minus signs do affect the value of $b_{{\bf m}_r}$ in Eq. (\ref{bmj}) since they are function of the squares of $Q_r$.

After having computed the amplitudes $b_{{\bf m}_r}$ from the vectors ${\bf m}_r$,  we need to sum over all of them according to
\be
s_4(n) =2\sum_{0\leq j_1<j_2<j_3<j_4\leq n-1} \left[ Q_{0}^{2\a}+ Q_{1}^{2\a}+ Q_{2}^{2\a}\right]\,,
\ee 
where we factorized out the factor $(4n^2)^{-2\a}$ according to the definition (\ref{Fexpintro}) 
and we used the positivity of the factors $Q_r$ (otherwise the absolute value of each $Q_r$ should be used).
The factor $2$ comes from the two possible vectors (i.e.~starting with $\pm1$) of each type.

This result for $s_4(n)$ agrees with the prediction of short-length expansion, using Eq. (\ref{d4})
and summing over $j_p$. The three $Q_r$ correspond to the three possible four-point correlation functions of   
the vertex operator $V_{\pm\a}(z)$, as clear by comparison. The factor two in front corresponds to the two sets of non vanishing
correlators.

It is also possible to further simplify $s_4(n)$. We can use the invariance under translation of the factor $Q_r$.
This implies that if we, for example fix $j_{2,3,4}$, the sum does not depend on $j_1$ that we can fix to $0$ 
and we get an overall factor    
a factor $n-{j_{4}}$, leading to 
\begin{eqnarray}\fl
s_4(n) &=& \sum_{1\leq j_2<j_3<j_4\leq n-1} 
2(n-j_{4}) \left[ Q_{0}^{2\a}+ Q_{1}^{2\a}+ Q_{2}^{2\a} \right]
\\ \fl
&=&
\frac{n}{2} 
\sum_{1\leq j_2<j_3<j_4\leq n-1} 
[Q_{0}^{2\a}+ Q_{1}^{2\a}+ Q_{2}^{2\a}]
=
\frac{n}{2} \sum_{1\leq j_2<j_3<j_4\leq n-1} 
[Q_{0}^{2\a}+ 2 Q_{1}^{2\a}]. 
\end{eqnarray}
In the last line we have used the identity 
\be
\sum_{1\leq j_2<j_3<j_4\leq n-1}  Q^{2\a}_1=
\sum_{1\leq j_2<j_3<j_4\leq n-1}  Q_2^{2\a}\,, \quad {\rm for\, any}\, \a\,,
\ee
that is valid only at the level of the sum (and after fixing $j_1=0$), but not for the single term.
This is a consequence of the permutation symmetry of the $Q_r$'s. 
Another interesting relation, that is particularly useful in the case $\a=1/4$ (and so for the Ising model in the following) is
\begin{equation}
\label{Q rel}
Q_{0}^{{1}/{2}} = Q_{1}^{{1}/{2}}+Q_{2}^{{1}/{2}}\,,
\end{equation}
which can be easily shown by using the prosthaphaeresis formula involving the product $\sin A \sin B$.

\subsection{Second order for the Ising model}

We can employ the results of the previous subsection 
to compute the second non vanishing term in the expansion of ${\cal F}_n(x)$ for the Ising model.  
For a given characteristic $(\bm\varepsilon,\bm\delta)$,  the vectors ${\bf m} \in \mathbf{Z}^{n-1}$ contributing to the $O(x^{1/2})$ order  
are characterized by $a_{\bf m+\bm\varepsilon}=1/2$.
From Eq. (\ref{deltaeff}), we have
\be
s_4(n)=\frac{(4n^2)^{1/2}}{2^{n-1}}  \sum_{\bf\bm\varepsilon\neq 0} \sum_{\bm\delta}\;
\Big|\sum_{{\bf m} \,{\rm s.t.}\,  a_{\bf m+\bm \varepsilon}=1/2} e^{2\pi i ({\bf m}+\bm\varepsilon)^{{\rm t}}\cdot \bm\delta} 
e^{b_{\bf m+\bm\varepsilon}}\Big|\,.
\ee
The vectors ${\bf m}$ contributing to this sum (i.e.~with $a_{\bf m+\bm\varepsilon}=1/2$) are such that ${\bf m}+\bm\varepsilon$ are half of the 
corresponding ones for the free boson, i.e.~half of ${\bf m}_r$ introduced in Eqs. (\ref{mvecdef}). 
Thus, as for the boson, the sum over ${\bf m}$ can be rewritten as sums over $j_{1,2,3,4}$, but the 
choice of ${\bf m}$ also fixes  $\bm\varepsilon$, and so this sum can be dropped.
Using the value of $b_{{\bf m}_r}$ in Eq. (\ref{bmj}) we have
\begin{eqnarray}\fl
s_4(n) &=& \sum_{0\leq j_1<j_2<j_3<j_4\leq n-1}
 \frac{1}{2^{n-1}} 
 \sum_{\bm\delta}\\ \fl
& &
2\left| 
Q_{0}^{1/2} \cos(\pi  {\bf m}_{0} \cdot \bm\delta)+
Q_{1}^{1/2} \cos(\pi  {\bf m}_{1} \cdot \bm\delta)+
{Q_{2}}^{1/2} \cos(\pi  {\bf m}_{2} \cdot \bm\delta)
\right|\,.
\nonumber
\end{eqnarray}
To write the sum over $\bm\delta$ in a manipulable way, we notice that the scalar products are very simple to write as
\be\fl
{\bf m}_{0} \cdot \bm\delta= \frac{q_1+q_2}2\,,\qquad
{\bf m}_{1} \cdot \bm\delta= \frac{q_1-q_2}2\,,\qquad
{\bf m}_{2} \cdot \bm\delta= \frac{q_1+q_2+2q_0}2\,,
\ee
where $q_1$ is the number of $1/2$ in $\bm\delta$ in the part $j_{21}$ of ${\bf m}_r$,
$q_2$ the number of $1/2$ in $\bm\delta$ in the part $j_{43}$ of ${\bf m}_r$,
and $q_0$ the number of $1/2$ in $\bm\delta$ in the part $j_{32}$.
The argument of the sum does not depend on the presence of $1/2$ in the $j$-th component of
$\bm\delta$ with $j>j_{4}$ and $j\leq j_1$, so that
the $2^{n-1-j_{41}}$ possible choices give equal contributions providing an overall factor $2^{n-1-j_{41}}$ in front of the sum over 
$q_{0,1,2}$. 
Putting everything together, we have 
\begin{eqnarray} \fl
s_4(n) &=&\sum_{0\leq j_1<j_2<j_3<j_4\leq n-1}
 \frac{1}{2^{j_{41}-1}}
 \sum_{q_1 =0}^{j_{21}} \sum_{q_2=0}^{j_{43}} \sum_{q_0 =0}^{j_{32}}
 {j_{21} \choose q_1}\, {j_{43} \choose q_2}  {j_{32} \choose q_0}\\ \fl
& 
 \times& 
\left| \cos \big[\frac{\pi}{2}(q_1+q_2)\big]
Q_{0}^{{1}/{2}}+
\cos \big[\frac{\pi}{2}(q_1-q_2)\big]
Q_{1}^{{1}/{2}}
+\cos \big[\frac{\pi}{2}(q_1+q_2+2q_0)\big]
Q_{2}^{{1}/{2}}
\right|. \nonumber
\end{eqnarray}
Since, for any triplet of integers $\{q_1,q_2,q_0\}$, the parity of $q_1+q_2$, $q_1-q_2=(q_1+q_2)-2q_2$ and $q_1+q_2+2q_0$ is the same, the terms having odd $q_1+q_2$ do not contribute. Thus, we have
\begin{eqnarray}
\label{F2 ising step2} \fl
s_4(n) &=& \sum_{0\leq j_1<j_2<j_3<j_4\leq n-1}
 \frac{1}{2^{j_{41}-1}}
 \sum_{q_1 =0}^{j_{21}} \sum_{q_2=0}^{j_{43}} \sum_{q_0 =0}^{j_{32}}
{j_{21} \choose q_1}\, {j_{43} \choose q_2}  {j_{32} \choose q_0}\\ \fl
& &
\label{F2 ising step2 bis}
\times 
\left|\, \cos \left(\frac{\pi}{2}(q_1+q_2)\right) \right|
\left|
Q_{0}^{{1}/{2}}+
(-1)^{q_2}
Q_{1}^{{1}/{2}}+
(-1)^{q_0}
Q_{2}^{{1}/{2}}
\right|\,,
\end{eqnarray}
where $|\cos(\pi (q_1+q_2)/2)|$ simply selects the terms with even $q_1+q_2$. Thus, the sums over $q$'s in (\ref{F2 ising step2}) can be organized as follows
\begin{eqnarray}\fl
\label{q sums step1}
 \sum_{q_1 }\sum_{q_2} \sum_{q_0} \bigg|_{{\rm even }\ q_1+q_2}
 &=&
  \bigg[ 
  \bigg(\sum_{{\rm even }\ q_1}+ \sum_{{\rm odd }\ q_1}\bigg)
  \bigg(\sum_{{\rm even }\ q_2}+ \sum_{{\rm odd }\ q_2}\bigg)
  \bigg] \bigg|_{{\rm even }\ q_1+q_2}\,
  \sum_{q_0} \hspace{1cm}\\ \fl
  &=&
    \bigg[ 
 \sum_{{\rm even }\ q_1}  \, \sum_{{\rm even }\ q_2}
 +  \sum_{{\rm odd }\ q_1}  \, \sum_{{\rm odd }\ q_2}
  \bigg]
  \bigg(\sum_{{\rm even }\ q_0}+ \sum_{{\rm odd }\ q_0}\bigg)\\ \fl
  &=&
  \label{q sums step3}
   \sum_{q_1,q_2,q_0\in 2\mathbf Z}
   +    \mathop{\sum_{q_1,q_2\in 2\mathbf Z}}_{q_0\in 2\mathbf {Z+1}}
   +    \mathop{\sum_{q_1,q_2\in 2\mathbf{Z+1}}}_{ q_0 \in 2\mathbf {Z}}
   + \sum_{q_1,q_2,q_0\in 2\mathbf {Z+1}}\,.
\end{eqnarray}
In Eq. (\ref{q sums step1}) the condition of even $q_1+q_2$ selects two of the four sums coming from $\sum_{q_1} \sum_{q_2}$.
Each of the four sums in (\ref{q sums step3}) is characterized by a specific sequence of relative signs between the three contributions to the general term in (\ref{F2 ising step2 bis}). Once the relative signs have been fixed, the dependence on $q$'s is only in the binomials. This means that we can perform the three sums over $q$'s  in each one of the four terms indicated in (\ref{q sums step3}). 
In order to make this sum, we find it convenient to leave the parity of three indices $q$'s unrestricted and introduce a factor 
$|\cos(\pi q/2)|$ or $|\cos(\pi (q+1)/2)|$ when $q$ is required to be respectively even or odd. Then, we have that
\begin{equation}
\label{q sum abs cos}
\sum_{q=0}^{j} {j \choose q} \left|\cos\left(\frac{\pi}{2}q\right)\right|
=
\sum_{q=0}^{j} {j \choose q} \left|\cos\left(\frac{\pi}{2}(q+1)\right)\right|
= 2^{j-1}\,.
\end{equation}
By using (\ref{q sums step3}) and (\ref{q sum abs cos}) we can perform the sums over $q$'s in (\ref{F2 ising step2}) and reduce them to the sum of four terms. Thus, we have
\begin{equation}\fl
\label{F2 ising step3}
s_4(n) = \sum_{0\leq j_1<j_2<j_3<j_4\leq n-1}
 \frac{1}{4}
 \sum_{\tau_1, \tau_2}
 \left|
Q_{0}^{{1}/{2}}+\tau_1
Q_{1}^{{1}/{2}}+\tau_2
Q_{2}^{{1}/{2}}
\right|\,,
\end{equation}
where $\tau_1 , \tau_2 \in \{-1,+1\}$. 
Now, we use Eq. (\ref{Q rel})
which allows us to further simplify the sum over $(\tau_1,\tau_2)$ in (\ref{F2 ising step3}), which becomes
\begin{eqnarray}\fl
 \sum_{\tau_1, \tau_2}
 \left|
Q_{0}^{{1}/{2}}+\tau_1
Q_{1}^{{1}/{2}}+\tau_2
Q_{2}^{{1}/{2}}
\right|  
= 2\left(
 \left| Q_{0}^{{1}/{2}}\right|
 + \left| Q_{1}^{{1}/{2}}\right|
 + \left| Q_{2}^{{1}/{2}}\right|
\right)
=4 Q_{0}^{{1}/{2}}\,,
\label{sumIsi}
\end{eqnarray}
where we have also used that $Q_{j}^{1/2} >0$.
Notice that, because of Eq. (\ref{Q rel}), the term corresponding to $(\tau_1,\tau_2)=(-,-)$ vanishes identically.
Thus, the final result is as simple as
\be
s_4(n) =\sum_{0\leq j_1<j_2<j_3<j_4\leq n-1}  Q_{0}^{{1}/{2}}\,,
\ee
that  is one quarter of the free-boson result for $\a=1/4$.
This value of $s_4(n)$ agrees with the result from short-length expansion using the four-point spin correlation function 
in Eq. (\ref{4ptIs}) and evaluating it at the roots of unit as in Eq. (\ref{d4}). This would give the sum of the three pieces in 
Eq. (\ref{sumIsi}), that we have shown how to further simplify. 

As for the free boson, we use translational invariance to fix $j_1=0$ at the price of the overall factor $n-j_4$, to rewrite $s_4(n)$ as 
\begin{eqnarray}\fl
\label{F2 ising final}
s_4(n) &=&\sum_{1\leq j_2<j_3<j_4\leq n-1}
(n-j_4) Q_{0}^{{1}/{2}}
=  \frac{n}{4}\sum_{1\leq j_2<j_3<j_4\leq n-1}
Q_{0}^{{1}/{2}}\,.
\end{eqnarray}

\section{The analytic continuation}
\label{ancont}

In this section we consider the analytic continuation of $s_2(n)$, 
and in particular we obtain its derivative at $n=1$ giving the small $x$ behavior of the entanglement entropy.
In Ref. \cite{cct-09} we provided a general method allowing a very precise numerical computation of $s_2(n)$ for non-integer $n$,
but we have been able to give a full analytic answer only for specific values of $\alpha$, i.e.~for  $\a=0,1/2,1$. 
This has been generalized in Ref. \cite{h-10} to any integer multiple of $\a/2$.
Here we fill this gap and give a full analytic answer for the analytic continuation for any value of $\a$. 

We start rewriting $s_2(n)$ as (we temporarily drop the factor ${\cal N}$)
\be
s_2(n)=\frac{n}2 \sum_{j=1}^{n-1} \frac1{\left[ \sin^2\left(\frac{2\pi j}{2n}\right) \right]^{\alpha}}\,,
\label{s2start}
\ee
and since it is periodic, let us write a Fourier series
\be
(\sin^2 (u/2))^{-\a}=\sum_{k\in \mathbf{Z}} f(k) e^{iku}\,,
\ee
where 
\be\fl
f(k)=\frac1{2\pi}\int_0^{2\pi} (\sin^2 (u/2))^{-\a}e^{-iku}du=\frac1\pi\int_0^\pi (\sin^2 y)^{-\a} e^{-2iky} dy\,.
\ee
For $\a<1/2$, this integral can be done (see e.g 3.631 of Gradshteyn and Ryzhik \cite{gr})
\be
f(k)=\frac{2^{2\a} e^{-i\pi k}\Gamma (1-2\a)}{\Gamma(1-\a+k)\Gamma(1-\a-k)}\,.
\ee
Note that $f(k) \sim |k|^{-1+2\a}$ as $|k|\to\infty$, as expected by power-counting.

We need now to perform the sum 
\be
\sum_{j=1}^{n-1} e^{i2\pi j k/n}\,,
\ee
which gives $n -1$ if $k = 0$ (mod $n$), and $-1$ otherwise. The sum for $s_2(n)$ can be then written as 
\be
s_2(n)= \frac12 \sum_{k\in \mathbf{Z}} [n f(n k)-f(k)]\,,
\label{s2wrong}
\ee
that vanishes at $n = 1$ as it must. This formula {\it could} provide the required analytic continuation, but it should be handle with 
a lot of care. Indeed there are two main problems with this sum, one  numerical and the other  conceptual. 

\begin{figure}
\includegraphics[width=0.7\textwidth]{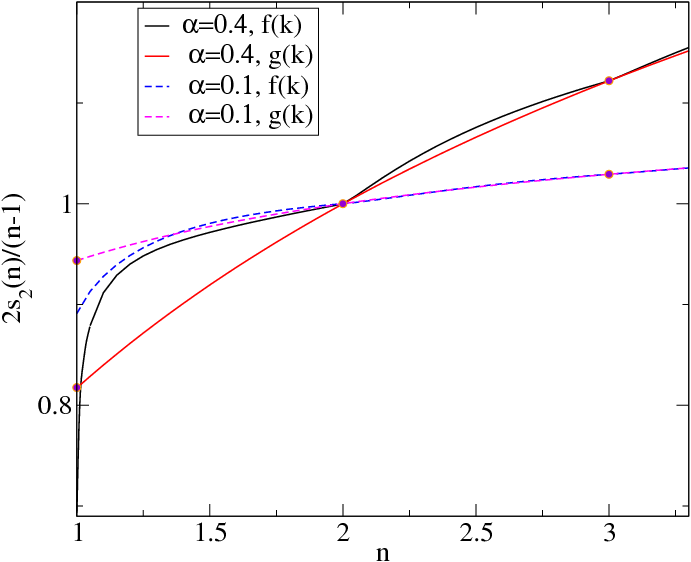}
\caption{This figure shows the importance of the Carlson's theorem to choose the correct analytic continuation.
For two values of $\a$ ($0.1$ and $0.4$), we report $s_n/(n-1)$ as function of $n$, as obtained by summing with 
the function $f(k)$ and $g(k)$  (i.e.~using Eq. (\ref{s2wrong}) or Eq. (\ref{s2final}) respectively).
It is evident that when using $f(k)$ an oscillatory behavior (with period $1$)  is superimposed to a smooth curve, 
while for $g(k)$ the resulting $s_2(n)$ is smooth. 
For integer numbers $\geq2$ they have the same value, as they should.  
The limits for $n\to1$ are very different and the one obtained from $g(k)$ is the correct one. 
}
\label{fig-carl}
\end{figure}

Let us start with the first problem. 
The two subsums defining $s_2(n)$ (i.e.~$\sum nf(nk)$ and $\sum f(k)$)  are both divergent. 
Their sum can be made convergent only if the two terms are summed in the proper order, i.e.~if 
we use the same upper limit for $nk$ in the first and $k$ in the second and only within this recipe we can take the difference and 
 send the cutoff to infinity.
If one makes the sum naively, it diverges. 
For integer $n$, by summing numerically in this way, we reproduce easily the correct results obtained by the direct sum 
Eq. (\ref{s2start}).  
However, for non-integer $n$ the obtained sum is not a smooth curve (see Fig. \ref{fig-carl}), signaling some 
troubles with this analytic continuation that leads to the second problem. 

Indeed, it is obvious that the analytic continuation of a function defined only on integer numbers is not unique, because, e.g.~we can 
multiply it by some function like $\cos^2(\pi n)$ (that is $1$ on the integer) or add any analytic function times $\sin (\pi n)$ that vanishes 
for integer $n$. However, all these analytic continuations with the same values on integer numbers have different behavior at 
$\pm i\infty$ and this is the essence of the Carlson's theorem \cite{carlson}, stating (in physical words) that the analytic continuation 
is unique once the desired behavior at $\pm i \infty$ is imposed.  
While at first sight this can seem a problem, it has been indeed shown that this arbitrariness 
can be a great advantage. For example, it can provide the solutions of different physical problems with a single result for integer numbers, 
as e.g.~done 
in Ref. \cite{cs-08} where correlation functions of one-dimensional models with different statistics have been obtained by the same 
replicated form, only requiring different behavior at $\pm i\infty$.

Going back to our problem, it should be clear that to select the correct analytic continuation we should have a look to $n\to\infty$, 
i.e.~we should consider the behavior of $f(k)$ for $k\to\infty$. Exploiting the asymptotic expansion of the $\Gamma$ function \cite{gr},
we have
\bea\fl
f(k\to\pm \infty)&\simeq& 2^{2\a} e^{-i \pi k} \Gamma (1-2\a)\frac{\sin[(\a+|k|)\pi]}{\pi |k|^{1-2\a}}\\
\fl&=&
2^{2\a} (\cos (\pi k)-i\sin(\pi k)) \Gamma (1-2\a)\frac{\sin(\a\pi)\cos(\pi k)+ \sin(|k|\pi)\cos(\pi \a) }{\pi |k|^{1-2\a}}
\,.\nonumber
\eea
 In this simple formula we see in action all the pieces for which Carlson's theorem has been formulated: 
there are additive $\sin(\pi k)$ pieces (that do not change the sum for integer $k$) and multiplicative $\cos^2\pi k$ (that 
just give $1$ on integers). Both factors change the value of $f(k)$ [and so of $s_2(n)$] only for non-integer values.
Carlson's theorem states that we need to impose the required analytic behavior for $\pm i\infty$ to select the correct analytic continuation.
In our case we require that $s_2(n)$ (and so any term in the sum $f(n k)$) does not blow up exponentially. 
This implies that all the $\sin (\pi k)$ terms are set to zero and the $\cos^2(\pi k)$ to one. 
In order to do so, we perform the replacement
\be
f(k\to\pm\infty)\to g(k\to\pm\infty)\equiv 
2^{2\a} \Gamma (1-2\a)\frac{\sin(\a\pi) }{\pi |k|^{1-2\a}}\,.
\ee
$f(k\to\infty)$ and $g(k\to\infty)$ have the same value for integer $k$, but they are different for non-integer ones. 

Being guided by the asymptotic behavior for large $k$, we can employ the same trick for the full $f(k)$ in order to have the correct
analytic continuation. We perform the replacement
\bea\fl 
f(k)
&\to&
 g(k)\equiv \frac{f(k) \sin(\pi\a)e^{i \pi k}}{\sin[\pi(\a+|k|)]}=
 \frac{2^{2\a}\Gamma (1-2\a)\sin(\pi\a)}{\sin[\pi(\a+|k|)] \Gamma(1-\a+k)\Gamma(1-\a-k)}\\ \fl
&=& \frac{4^\a \sin (\pi  \a) \Gamma (1-2 \a) \sin[\pi  (\a-|k|)] \Gamma (\a-k) \Gamma (\a+k)}{\pi ^2}\,.
\eea
Again, $f(k)$ and $g(k)$ have the same values for integer $k$, but it is easy to check that while $f(k)$ oscillates  for 
non integer $k$ (signaling `bad exponential'  behavior at $\pm i\infty$), $g(k)$ is monotonic.

The sum for $s_2(n)$ can be rewritten as 
\be\fl
s_2(n)=\frac12
\sum_{k=-\infty}^\infty [n g(n k)-g(k)]
=\frac{n-1}2 g(0)+
\sum_{k=1}^\infty [n g(n k)-g(k)]\,,
\label{s2final}
\ee
that is the {\it analytic continuation we were searching for}, but valid only for $\a<1/2$.
As stressed above,  this sum is convergent if for the first and second piece that same cutoff is used and after sent to $\infty$.
We performed numerically this sum for non-integer $n$ 
finding perfect agreement with the results in Ref. \cite{cct-09} obtained by Borel resummation. 
Fig. \ref{fig-carl} shows the difference in the summation of $s_2(n)$ obtained for non-integer $n$ while using the function 
$g(k)$ instead of $f(k)$.

Eq. (\ref{s2final}) allows analytic manipulations that give the exact result  for $n=1$. 
Let us use the integral representation of the product of $\Gamma$ functions
\be
\Gamma(p)\Gamma(q)= 2\Gamma(p+q)\int_0^{\pi/2} (\cos\phi)^{2p-1}(\sin\phi)^{2q-1}\,,
\ee
to rewrite
\bea\fl
 g(k)&=&\frac{4^\a \sin (\pi  \a) \Gamma (1-2 \a)}{\pi ^2} \sin [\pi  (\a-|k|)]
2\Gamma(2 \a)\int_0^{\pi/2} \hspace{-3mm} (\cos\phi)^{2\a-2k-1}(\sin\phi)^{2\a+2k-1} d\phi \nonumber\\ \fl
&=&\frac{2^{2 \a} \sec (\pi  \a)}{\pi }\sin [\pi  (\a-|k|)]
\int_0^{\pi/2} \hspace{-1mm} \big(\frac{\sin2\phi}2\big)^{2\a-1} (\tan\phi)^{2k} d\phi\,,
\eea
where we used also  $4^\a \Gamma(1 - 2 \a) \sin(\a \pi) 2 \Gamma (2 \a)/\pi^2  =2^{2 \a} \sec (\pi  \a)/{\pi }$.

In order to evaluate $\sum_{k=1}^\infty n g(n k)$, we exchange the order of sum and integration. 
Calling $v=\tan\phi$, we need  the sum
\bea\fl
I_n&=&\sum_{k=1}^\infty  \sin [\pi  (\a-n k)] v^{2k n}= 
\frac{v^{2 n} \left[\sin (\pi  (\a-n))-\sin (\pi  \a) v^{2 n}\right]}{v^{4 n}-2 \cos (\pi  n) v^{2 n}+1}\,,
\eea
that is strictly valid only for $|v|<1$ (i.e.~$\phi<\pi/4$), but we will assume its validity everywhere.
It is easy to take the derivative of $n I_n$ wrt $n$ and set $n=1$, obtaining
\bea\fl
\left. \frac{\partial  (nI_n)}{\partial n}\right|_{n=1}&
=&\frac{v^2}{(1+v^2)^2}[\pi \cos(\pi\a)-(1+v^2+2\log v)\sin(\pi\a)]
\nonumber \\ \fl &=&
\sin^2(\phi ) \cos ^2(\phi ) \left[\pi \cos(\pi  \a) -\sin (\pi \a) \left[ \sec ^2\phi +2 \log (\tan \phi)\right]\right].
\eea

Plugging the various pieces together we have\footnote{
The following integrals are useful for this calculation:
\bea
\int_0^{\pi/2} (\sin 2\phi )^{2 \a+1}= \frac{\sqrt{\pi } \Gamma (\a+1)}{2\Gamma \left(\a+\frac{3}{2}\right)}\,, 
\nonumber \\
\int_0^{\pi/2}  (\sin 2\phi)^{2 \a+1} (2 \log\tan\phi + \sec^2\phi)=  \frac{\sqrt\pi \Gamma (\a)}{\Gamma \left(\a+\frac{1}{2}\right)}=
\frac{4^\a \Gamma (1-2 \a)}{\Gamma (1-\a)^2}\frac{\pi}{\tan \pi\a}\,. \nonumber
\eea} 
\bea\fl
2 s'_2(1)&=&g(0)+\frac{1}{\pi  \cos (\pi  \a)} 
\left[ \int_0^{\pi/2} (\sin 2\phi)^{2 \a+1} \pi  \cos (\pi  \a)
\right. \nonumber\\ \fl&&\qquad \left.
-\int_0^{\pi/2} (\sin 2\phi )^{2 \a+1}
\sin (\pi  \a) (\sec ^2\phi+2 \log (\tan \phi))\right] \nonumber\\ \fl&
=&\frac{4^\a \Gamma (1-2 \a)}{\Gamma (1-\a)^2}+\frac{\sqrt{\pi } \Gamma (\a+1)}{2 \Gamma \left(\a+\frac{3}{2}\right)}
-\frac{\tan (\pi  \a) \Gamma (\a)}{\sqrt{\pi } \Gamma \left(\a+\frac{1}{2}\right)}=
\frac{\sqrt{\pi } \Gamma (\a+1)}{2 \Gamma \left(\a+\frac{3}{2}\right)}\,,
\eea
that is the result anticipated in Eq. (\ref{s21fin}), where we restored the multiplicative factor ${\cal N}$.
While this result has been derived for $\a<1/2$, the final form of $s'_2(1)$ is analytic all the way up to $\alpha=\infty$ and indeed it 
agrees both with the numerical results of Ref. \cite{cct-09} for all $\a<1$ and it is straightforward to show that for positive 
integer $\a$ reproduces the right result (in this case, $s_2(n)$ is a polynomial  \cite{h-10} and it is easy to find $s_2'(1)$). 

\subsection{Consequences for the Ising model}

In the case of the Ising model, the first order in $x$ of the scaling function ${\cal F}_n(x)$ is half of the result for free boson at $\a=1/4$ 
(cf. Eq. (\ref{F1 ising final})). Thus the scaling function for the entanglement entropy is 
\be\fl
{\cal F}_{VN}(x)=\frac{\partial}{\partial n} {\cal F}_n (x)\Big|_{n=1}=
 \frac{\sqrt{\pi } \Gamma \left(\frac{5}{4}\right)}{4\sqrt2  \Gamma\left(\frac{7}{4}\right)} x^{1/4}
   +O(x^{1/2})= 0.309012\dots x^{1/4}+\cdots
\,.
\ee
Note the numerical value of the amplitude of the term $x^{1/4}$ is close to (but definitively different than) $1/\pi= 0.318\dots$, 
that was proposed in Ref. \cite{atc-09} on the basis of purely numerical results. 
We checked that the numerical precision of the data in Ref. \cite{atc-09} does not allow to distinguish between these too close values. 


\section*{Acknowledgments}

We thank B. Doyon, M. Fagotti,  M. Headrick, and A. Scardicchio for very useful discussions. 
ET thanks the Physics department of the University of Pisa for the hospitality.
JC's research was supported in part by the EPSRC under Grant EP/D050952/1 and by the National Science Foundation under Grant NSF PHY05-51164. He acknowledges the facilities of the KITP Santa Barbara where this paper was completed.
This work has been partly done when PC and ET were guests of  the Galileo Galilei Institute in Florence whose hospitality is kindly 
acknowledged.
ET is mainly supported by Istituto Nazionale di Fisica Nucleare (INFN) through a Bruno Rossi fellowship and also by 
funds of the U.S. Department of Energy (DoE) under the cooperative research agreement DE-FG02-05ER41360.

\appendix

\section{Mapping of two intervals to the cylinder geometry}

Some of the discussion of the main part of the paper may be more
readily comprehended if we conformally map the $n$-sheeted
conifold consisting of $n$ planes connected along the two
intervals, to another consisting of $n$ open cylinders connected
in a certain way along their top and bottom edges.

\begin{figure}
\includegraphics[width=0.5\textwidth]{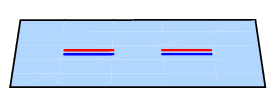}
\includegraphics[width=0.65\textwidth]{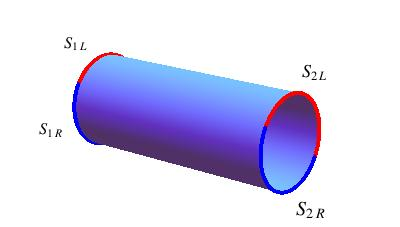}
\caption{The cut plane represented on the left can be mapped through a conformal transformation 
to a cylinder of length $L$ and circumference $W$ (right) and so modular parameter $q=e^{-2\pi L/W}$.
The upper and lower edges of the cuts are mapped into
semicircular arcs $(S_{1L},S_{1R})$ and $(S_{2L},S_{2R})$ at each
end of the open cylinder.}
\label{planetocyl}
\end{figure}

\begin{figure}[b]
\centerline{\includegraphics[width=0.52\textwidth]{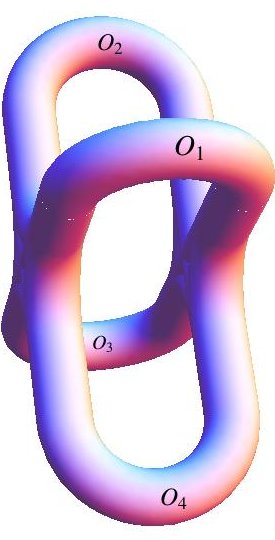}}
\caption{The equivalent of the $n$-sheeted Riemann surface (with $n=4$) for the cylinder geometry.  
$O_j$ represents the operator ``propagating'' in the cylinder $j$.}
\label{cage}
\end{figure}

For a single plane, cut open along the upper and lower edges of
the intervals $(u_1,v_1)$ and $(u_2,v_2)$ there is a conformal
mapping (whose explicit expression is not needed)
to a cylinder of circumference $W$ and length $L$. 
We introduce the modular parameter $q=e^{-2 \pi
L/W}$ which is a known function of the cross-ratio $x$ of the
original four points in the plane. The limit $x\to0$ corresponds
to $q\to0$. The upper and lower edges of the cuts are mapped into
semicircular arcs $(S_{1L},S_{1R})$ and $(S_{2L},S_{2R})$ at each
end of the open cylinder. This is illustrated in Fig.~\ref{planetocyl}. We
do this for each plane, labelled by $j\in[1,n]$. These are then
connected cyclically by their edges so that $S_{1R}^{(j)}$ is
connected to $S_{1L}^{(j+1)}$ and $S_{2R}^{(j)}$ is connected to
$S_{2L}^{(j+1)}$. The case $n=4$ is shown in Fig.~\ref{cage}.

Note, however, that this in general introduces new conical
singularities at the points where the semicircles meet. The total
angle subtended by a curve which encloses a singularity is now
$\pi n$, rather than $2\pi n$ as in the original geometry. This
changes the value of the trace anomaly at each singularity from
$(c/12)\big(n-(1/n)\big)$ to $(c/12)\big((n/2)-(2/n)\big)$. The
partition function in the coupled cylinder geometry thus has the
form
\be
Z_n^{\rm cylinder}\propto W^{-(c/3)\big((n/2)-(2/n)\big)}\,{\cal
F}^{\rm cyl}(q)\,.
\ee
However, the non-trivial dependence on $q$ should be conformally
invariant, that is
\be
{\cal F}_n^{\rm cyl}\big(q(x)\big)={\cal F}_n(x)\,,
\ee
where ${\cal F}_n(x)$ is the same function as in (\ref{Fn}). We
assume they are both normalized so that ${\cal F}_n^{\rm
cyl}(0)={\cal F}_n(0)=1$.

Note in particular that for $n=2$ there is no trace anomaly. This
because in this case, one of the cylinders can be reflected
$L\leftrightarrow R$ so that $S_{1L}^{(1)}$ is now sewn onto
$S_{1L}^{(2)}$ and $S_{2L}^{(1)}$ is sewn onto $S_{2L}^{(2)}$. For
$n=2$ we can also reflect the second cylinder in the plane of one
of its ends. This gives a cylinder of length $2L$ with its
opposite ends identified, that is a torus, with modular parameter
$e^{-4\pi L/W}=q^2$. As is well-known \cite{Cardy86} the torus
partition function encodes all the scaling dimensions of the
theory, that is
\begin{equation}\label{eq:F2}
{\cal F}_2^{\rm cyl}(q)=1+\sum_{k\not=0}(q^2)^{\Delta_k+\Db_k}\,.
\end{equation}
This is consistent with our result above that, for $n=2$, only
terms involving the 2-point functions of primary operators can
arise in the small $\ell$ expansion of ${\cal F}_2$, which has the
form
\be
{\cal F}_2\propto\sum_{\rm
primaries}d_k(\ell_1\ell_2/r^2)^{2(\Delta_k+\Db_k)}+\cdots\,,
\ee
where $d_k=\sum_{j\not=j'}d_{k;jj'}^2$ and the neglected terms are
the contributions of descendant fields. Presumably the full
identity of the above two expressions involves a complicated
connection between these neglected terms and the higher order
terms of the expansion of $q(x)$ in powers of $x$.

It is interesting to consider computing $Z_n^{\rm cylinder}$ for
general $n$ as a similar expansion in powers of $q$. This can done
by considering the infinitesimal generator $\hat H$ of
translations along each cylinder. As is well known \cite{Cardy86},
the eigenstates $|k\rangle$ of $\hat H$ are in 1-1 correspondence
with the scaling fields of the CFT, with corresponding eigenvalue
$(2\pi/W)(\Delta_k+\Db_k)-(\pi c/6W)$. The last term drops out if
we normalize so that ${\cal F}^{\rm cyl}_n(q=0)=1$. Now divide the Hilbert
space into two subspaces ${\cal H}_L$ and ${\cal H}_R$
corresponding to the $L$ and $R$ halves of the cylinder. Then each
state admits a Schmidt decomposition
\be
|k\rangle=\sum_{\alpha}
c_{k\alpha}|k;\alpha\rangle_L\otimes|\widetilde{k;\alpha}\rangle_R\,.
\ee
Note the parity operator $\hat P$ which reflects $L\leftrightarrow
R$ (and corresponds to complex conjugation in the planar geometry)
gives an isomorphism between ${\cal H}_L$ and ${\cal H}_R$.
However, the states $|k;\alpha\rangle_L$ and
$|\widetilde{k;\alpha}\rangle_R$ generated by the Schmidt
procedure are not in general mapped into each other by this
isomorphism.

Since the states $|k\rangle$ form an orthonormal basis,
\begin{equation}\label{eq:ortho}
\sum_{\alpha\alpha'}c_{k\alpha}c_{k'\alpha'}\,{}_L\langle
k;\alpha|k';\alpha'\rangle_L\,{}_R\langle
\widetilde{k;\alpha}|\widetilde{k';\alpha'}\rangle_R=
\delta_{kk'}\,.
\end{equation}
For $k=k'$ this follows from the orthonormality of the Schmidt
states and the fact that $\sum_\alpha c_{k\alpha}^2=1$, but in
general it is not true that $|k,\alpha\rangle$ and
$|k',\alpha'\rangle$ are orthogonal for $k\not=k'$.

We make this decomposition for each end of each cylinder, labelled
by $j$. The sewing procedure is then equivalent to taking inner
products of states in the $L$ space on cylinder $j$ with those in
the $R$ space of cylinder $j+1$. Explicitly
\begin{equation}\fl\label{eq:Fcyl}
{\cal F}^{\rm cyl}_n(q)\propto\sum_{\{k_j\}}
\left(\sum_{\{\alpha_j\}}\prod_{j=1}^nc_{k_j\alpha_j}\,\langle
k_j;\alpha_j|\widetilde{k_{j+1};\alpha_{j+1}}\rangle
\right)^2\,q^{\sum_j(\Delta_{k_j}+\Db_{k_j})}\,,
\end{equation}
where we have dropped the distinction between ${\cal H}_L$ and
${\cal H}_R$, assuming that these are identified using the
isomorphism under $\hat P$. However, when $n$ is even we can
relabel the basis for each even $j$ so that
$|k_j;\alpha_j\rangle\leftrightarrow|\widetilde{k_j;\alpha_j}\rangle$
and $\Delta_k\leftrightarrow\Db_k$ (equivalent to reflecting
cylinders with $j$ even). The expression inside the modulus sign
then becomes
\be\fl
\sum_{\{\alpha_j\}}\left(\prod_{j\ {\rm
odd}}^nc_{k_j\alpha_j}\,\langle
k_j;\alpha_j|k_{j+1};\alpha_{j+1}\rangle\right)\left(\prod_{j\
{\rm even}}^nc_{k_j\alpha_j}\,\langle
\widetilde{k_j;\alpha_j}|\widetilde{k_{j+1};\alpha_{j+1}}\rangle\right)\,.
\ee
(For odd $n$ we can do this with $(n-1)/2$ of the cylinders,
leaving a single factor of $\langle
k_j;\alpha_j|\widetilde{k_{j+1};\alpha_{j+1}}\rangle$.)

For $n=2$ we then see that the sum between the modulus sign has
the same form as that in (\ref{eq:ortho}) and therefore we recover
the torus partition function (\ref{eq:F2}).

It is illuminating to write (\ref{eq:Fcyl})  in another way.
Introduce the operators ${\hat M}_k:{\cal H}_L\to{\cal H}_L$:
\be
{\hat M}_k\equiv\sum_\alpha
c_{k\alpha}|\widetilde{k;\alpha}\rangle\langle k;\alpha|\,.
\ee
Then the expression in parentheses in (\ref{eq:Fcyl}) is simply
\be
{\rm Tr}_{{\cal H}_L}\,\prod_{j=1}^n{\hat M}_{k_j}\,.
\ee
Note that under the parity operation ${\hat M}\leftrightarrow
{\hat M}^\dag$. The fact that we can reflect an arbitrary number
of the cylinders without changing the partition function means
that we can, in the above matrix product, change ${\hat
M}_{k_j}\to {\hat M}_{k_j}^\dag$ in any number of the factors, up
to a sign. In fact the sign is determined by the parities of the
states $|k_j\rangle$: since $\hat P$ commutes with $\hat H$, the
eigenstates $|k\rangle$ can be chosen also to be eigenstates of
$\hat P$ with eigenvalues $(-1)^{\Delta_{k_j}-\Db_{k_j}}$. Note
however that we should make the same set of replacements ${\hat
M}\to {\hat M}^\dag$ for all $\{k_j\}$.

However, we can make a stronger statement, based on the
observation that we can consider the modulus $q_j$ of the $j$th
cylinder to depend on $j$, and also to be complex, corresponding
to twisting the cylinder before the sewing procedure. This no
longer corresponds to computing the R\'enyi entropy, but still
makes mathematical sense. In this case the $q$-dependence in
(\ref{eq:Fcyl}) is generalized to the form $\prod_j
q_j^{\Delta_j}{\bar q}_j^{\Db_j}$. Therefore the identity between
the different expressions obtained by reflecting each cylinder
independently must also hold for each set $\{k_j\}$ with the same
values of $\{\Delta_{k_j}\}$ and $\{\Db_{k_j}\}$. The consequence
of this is most easily seen in the case when all the $\{k_j\}$
correspond to primary operators. Then, barring accidental
degeneracies,
\be
{\rm Tr}\,\left({\hat M}_{k_1}{\hat M}_{k_2}\cdots {\hat
M}_{k_n}\right)\,,
\ee
is invariant, up to a sign determined by the overall parity, on
replacing any subset of the ${\hat M}_{k_j}$ by their hermitian
conjugates. Since this must be true for arbitrary $n$, the
simplest solution is that ${\hat M}_{k}^\dag=\pm {\hat M}_{k}$. In
the more general case, there is a finite-dimensional space of
descendent operators with the same values of $(\Delta_k,\Db_k)$,
and we have
\be
{\hat M}_{k}^\dag=\sum_{k'}A_{kk'}{\hat M}_{k'}\,,
\ee
where the sum is restricted to operators within this space, and
$A^*A=1$.

We can express the results of Sec.~\ref{Secope} in this formalism. The
case $n=1$ corresponds to a sphere, which has no modulus and
therefore there should be no $q$-dependence. This implies that
${\rm Tr}\,{\hat M}_k\propto \delta_{k0}$. Similarly, for $n=2$,
we have ${\rm Tr}\,{\hat M}_k{\hat M}_{k'}\propto\delta_{kk'}$.

If we take all the $k_j=0$ we get the leading term in the
partition function as $q\to0$:
\be
{\rm Tr}\,{\hat M}_0^n\propto W^{-(c/3)\big((n/2)-(2/n)\big)}\,.
\ee
For even $n$, we can also write the left side as
\be
{\rm Tr}\,\big({\hat M}_0{\hat M}_0^\dag)^{n/2}=\sum_\alpha
c_{0\alpha}^n\,.
\ee
For odd $n$ we get ${\rm Tr}\,\big(C_0^{n-1}{\hat M}_0\big)$,
where $C_0$ is a diagonal matrix with entries $c_{0\alpha}$. This
gives information about the spectrum of ${\hat M}_0$ and, since
this is valid for all $n$, suggests that its non-zero eigenvalues
are simply $c_{0k}$. However, since ${\cal H}_L$ is
infinite-dimensional, this may not hold in a strong sense.

Once this overall $W$ dependence is factored out, we see, that, at
least for primary operators
\be
{\rm Tr}\prod_{j=1}^n{\hat
M}_{k_j}\propto\langle\prod_j\phi_{k_j}(e^{2\pi ij/n})\rangle_{\bf C}\,,
\label{last}
\ee
where the rhs is the $n$-point function at the $n$th roots of
unity in the plane. It would be interesting to explore the
properties of the operators ${\hat M}_k$ further, as they appear
to encode the entanglement properties of the theory in a slightly
different way.
The operators $\phi_{k_j}$ in Eq. (\ref{last}) can be thought as ``propagating'' in the cylinders as pictorially shown in 
Fig. \ref{cage} (where they have been called $O_j$).


\section*{References}

\begin{thebibliography}{99}


\bibitem{cct-09}
P Calabrese, J Cardy, and E Tonni,
Entanglement entropy of two disjoint intervals in conformal field theory,
J. Stat. Mech. (2009) P11001.
  
  
\bibitem{cc-04}
P Calabrese and J Cardy,
Entanglement entropy and quantum field theory,
J. Stat. Mech. P06002 (2004).

\bibitem{cc-rev}
P Calabrese and J Cardy,
Entanglement entropy and conformal field theory,
J. Phys. A {\bf 42}, 504005 (2009).


\bibitem{revs} L Amico, R Fazio, A Osterloh, and V Vedral,
Entanglement in many-body systems, Rev. Mod. Phys. {\bf 80}, 517 (2008);
J Eisert, M Cramer, and M B Plenio, Area laws for the entanglement
entropy - a review, Rev. Mod. Phys. {\bf 82}, 277 (2010);
Entanglement entropy in extended systems,
P Calabrese, J Cardy, and B Doyon Eds, J. Phys. A {\bf 42} 500301 (2009).

\bibitem{Holzhey} C Holzhey, F Larsen, and F Wilczek,
Geometric and renormalized entropy in conformal field theory,
Nucl. Phys. B {\bf 424}, 443 (1994).

\bibitem{Vidal}
G. Vidal, J. I. Latorre, E. Rico, and A. Kitaev,
Entanglement in quantum critical phenomena,
Phys. Rev. Lett. {\bf 90}, 227902 (2003);
J. I. Latorre, E. Rico, and G. Vidal,
Ground state entanglement in quantum spin chains,
Quant. Inf. and Comp. {\bf 4}, 048 (2004).


\bibitem{cl-08}
P. Calabrese and A. Lefevre, 
Entanglement spectrum in one-dimensional systems,
Phys. Rev. A {\bf 78}, 032329 (2008).

\bibitem{fps-09}
S Furukawa, V Pasquier, and J Shiraishi,
Mutual information and compactification radius in a $c=1$ critical phase in 
one dimension, Phys. Rev. Lett. {\bf 102}, 170602 (2009).

\bibitem{cg-08}
M Caraglio and F Gliozzi, Entanglement entropy and twist fields,
JHEP 0811: 076 (2008)


\bibitem{ch-04}
H Casini and M Huerta,
A finite entanglement entropy and the c-theorem,
Phys. Lett. B {\bf 600} (2004) 142;
H Casini, C D Fosco, and M Huerta,
Entanglement and alpha entropies for a massive Dirac field in two dimensions,
J. Stat. Mech. P05007 (2005);
H Casini and M Huerta,
Remarks on the entanglement entropy for disconnected regions,
JHEP 0903: 048 (2009);
H Casini and M Huerta,
Reduced density matrix and internal dynamics for multicomponent regions, 
Class. Quant. Grav. {\bf 26}, 185005 (2009);
H Casini, Entropy inequalities from reflection positivity,
J. Stat. Mech. (2010) P08019.

\bibitem{ffip-08}
P Facchi, G Florio, C Invernizzi, and S Pascazio,
Entanglement of two blocks of spins in the critical Ising model,
Phys. Rev. A {\bf 78}, 052302 (2008).

\bibitem{kl-08} I Klich and L Levitov,
Quantum noise as an entanglement meter,
Phys. Rev. Lett. {\bf 102}, 100502 (2009).

\bibitem{rt-06}
S. Ryu and T. Takayanagi, 
Holographic derivation of entanglement entropy from AdS/CFT,
Phys. Rev. Lett. {\bf 96} (2006) 181602;
S. Ryu and T. Takayanagi, 
Aspects of holographic entanglement entropy,
JHEP 0608: 045 (2006);
M Headrick and T Takayanagi,
A holographic proof of the strong subadditivity of entanglement entropy,
Phys. Rev. D {\bf 76}, 106013 (2007);
T. Nishioka, S. Ryu, and T. Takayanagi, 
Holographic entanglement entropy: an overview,
J. Phys. A {\bf 42} (2009) 504008.

\bibitem{hr-08}
V E Hubeny and M Rangamani,
Holographic entanglement entropy for disconnected regions,
JHEP 0803: 006 (2008).

\bibitem{atc-09}
V Alba, L Tagliacozzo, and P Calabrese,
Entanglement entropy of two disjoint blocks in critical Ising models, 
Phys. Rev. B {\bf 81} (2010) 060411.

\bibitem{ip-09}
F Igloi and I Peschel, On reduced density matrices for disjoint subsystems, 
2010 EPL {\bf 89} 40001.

\bibitem{fc-10}
M Fagotti and P Calabrese,
Entanglement entropy of two disjoint blocks in XY chains,
J. Stat. Mech. (2010) P04016.


\bibitem{h-10}
M Headrick, Entanglement R\'enyi entropies in holographic theories, 1006.0047.

\bibitem{fc-10b}
M Fagotti and P Calabrese,
Universal parity effects in the entanglement entropy of XX chains with open boundary conditions,
1010.5796.

\bibitem{t-10}
E Tonni,
Holographic entanglement entropy: near horizon geometry and disconnected regions, 1011.0166.

\bibitem{c-10}
P Calabrese,
Entanglement entropy in conformal field theory: New results for disconnected regions,
J. Stat. Mech. (2010) P09013.

\bibitem{Neg}
H. Wichterich, J. Molina-Vilaplana, and S. Bose,
Scale invariant entanglement at quantum phase transitions,
Phys. Rev. A {\bf 80}, 010304(R) (2009);
S. Marcovitch, A. Retzker, M. B. Plenio, and B. Reznik,
Critical and noncritical long range entanglement in the Klein-Gordon field,
Phys. Rev. A {\bf 80}, 012325 (2009);
H. Wichterich, J. Vidal, and S. Bose,
Universality of the negativity in the Lipkin-Meshkov-Glick model, Phys. Rev. A {\bf 81}, 032311 (2010).

\bibitem{bpz-84}  A.~A.~Belavin, A.~M.~Polyakov and A.~B.~Zamolodchikov,
Infinite conformal symmetry in two-dimensional quantum field theory,
1984 Nucl. Phys. B {\bf 241} 333.

\bibitem{book}
P Ginsparg, Applied conformal field theory, in: {\it Les Houches, session XLIX (1988), Fields, strings and critical phenomena}, 
Eds. E. Br\'ezin and J. Zinn-Justin, Elsevier, New York (1989).

  
\bibitem{Dijkgraaf:1987vp}
R Dijkgraaf, E P Verlinde and H L Verlinde,
$c=1$ Conformal Field Theories on Riemann Surfaces, Commun.\ Math.\ Phys.\  {\bf 115} (1988) 649.

\bibitem{orb2}
L Alvarez-Gaume, J B Bost, G W Moore, P C Nelson and C Vafa,
Bosonization on higher genus Riemann surfaces,
Commun. Math. Phys.  {\bf 112} (1987) 503;
D Bernard, $Z_2$-twisted fields and bosonization on Riemann surfaces,
Nucl. Phys. B {\bf 302} (1988) 251.

\bibitem{Dixon:1986qv}
L J Dixon, D Friedan, E J Martinec and S H Shenker,
The conformal field theory of orbifolds,
Nucl. Phys.  B {\bf 282} (1987) 13.

\bibitem{z-87}
Al B Zamolodchicov, Conformal scalar field on the hyperelliptic curve 
and critical Ashkin-Teller multipoint correlation functions,
Nucl. Phys. B {\bf 285} (1987) 481;
V. G. Knizhnik, Analytic fields on Riemann surfaces. II,
Communn. Math. Phys. {\bf 112}, 567 (1987). 
 

\bibitem{gr}
I S Gradshteyn and I M Ryzhik, Table of Integrals, Series, and Products (Academic, London, 1980).

\bibitem{carlson}
L A Rubel, Necessary and Sufficient Conditions for Carlson's Theorem on Entire Functions, 
Proc. Natl. Acad. Sci. USA  {\bf 41},  601 (1955).

\bibitem{cs-08}
P Calabrese and R Santachiara, Off-diagonal correlations in one-dimensional anyonic models: A replica approach, 
J. Stat. Mech. P03002, 2009.

\bibitem{Cardy86}
J L Cardy,  Operator Content of Two-Dimensional Conformally Invariant Theories, 
Nucl. Phys. B {\bf 270}, 186 (1986).

\end{thebibliography}
\end{document}